\documentstyle[aps,pra,psfig,multicol]{revtex}

\begin{document}
\draft
\title{Estimates for practical quantum cryptography}
\author{Norbert L\"utkenhaus}
\address{Helsinki Institute of Physics, PL 9, FIN-00014
Helsingin yliopisto, Finland}
\date{\today}
\maketitle
\begin{abstract}
In this article I present a protocol for quantum cryptography which is
secure against attacks on individual signals. It is based on the
Bennett-Brassard protocol of 1984 (BB84). The security proof is
complete as far as the use of single photons as signal states is
concerned. Emphasis is given to the practicability of the resulting
protocol.  For each run of the quantum key distribution the security
statement gives the probability of a successful key generation and the
probability for an eavesdropper's knowledge, measured as change in
Shannon entropy, to be below a specified maximal value.
\end{abstract}
%\narrowtext
%\begin{multicols}{2}
\section{Introduction}
Quantum Cryptography is a technique for generating and distributing
cryptographic keys in which the secrecy of the keys is guaranteed by
quantum mechanics. The first such scheme was proposed by Bennett and
Brassard in 1984 (BB84 protocol) \cite{bennett84a}. Sender and
receiver (conventionally called Alice and Bob) use a quantum channel,
which is governed by the laws of quantum mechanics, and a classical
channel which is postulated to have the property that any classical
message sent will be faithfully received. The classical channel will
also transmit faithfully a copy of the message to any eavesdropper,
Eve. Along the quantum channel a sequence of signals is sent chosen at
random from two pairs of orthogonal quantum states. Each such pair
spans the same Hilbert space. For example, the signals can be realized
as polarized photons: one pair uses horizontal and vertical linear
polarization ($+$) while the other uses linear polarization rotated by
$45$ degrees ($\times$).  Bob at random one of two measurements each
performing projection measurements on  the basis $+$ or
$\times$.  The {\it sifted key} \cite{huttner94a} consists of the
subset of signals where the bases of signal and measurement coincide
leading to deterministic results. This subset can be found by exchange
of classical information without revealing the signals
themselves. Any  attempt of an eavesdropper to obtain information
about the signals leads to a non-zero expected error rate in the
sifted key and makes it likely that Alice and Bob can detect the
presence of the eavesdropper by comparing a subset of the sifted key
over the public channel. If Alice and Bob find no errors they conclude
(within the statistical bounds of error detection) that no
eavesdropper was active. They then translate the sifted key into a sequence
of zeros and ones which  can be used, for example, as a one-time pad
in secure communication.

Several quantum cryptography experiments have been performed. In the
experimental set-up noise is always present leading to a bit error
rate of, typically, 1 to 5 percent errors in the sifted key
\cite{marand95a,zbinden98a,franson94a,buttler98a}. Alice and Bob can
not even in principle distinguish between a noisy quantum channel and
the signature of an eavesdropper activity. The protocol of the key
distribution has therefore to be amended by two steps. The first is
the {\it reconciliation} (or error correction) step leading to a key shared
by Alice and Bob. The second step deals with the situation that the
eavesdropper now has to be assumed to be in the possession of at least
some knowledge about the reconciled string. For example, if one
collects some parity bits of randomly chosen subsets of the reconciled
string as a new key then the Shannon information of an eavesdropper
on that new, shorter key can be brought arbitrarily close to zero
by control of the number of parity bits contributing towards it. This
technique is the generalized privacy amplification procedure by
Bennett, Brassard, Cr\'epeau, and Maurer \cite{bennett95a}.

The final measure of knowledge about the key used in this article is
that of change of Shannon entropy. If we assign to each potential key
$x$ an a-priori probability $p(x)$ then the Shannon entropy of this
distribution is defined as
\begin{equation}
S\left[ p(x)\right] = - \sum_x p(x) \log p(x) \; .
\end{equation}
Note that all logarithms in this article refer to basis $2$. The
knowledge Eve obtains on the key may be denoted by $k$ and leads to an
a-posteriori probability distribution $p(x|k)$. The difference between
the Shannon entropy of the a-priori and the a-posteriori probability
distribution is a good measure of Eve's knowledge:
\begin{equation}
\Delta_S(k) = S\left[ p(x)\right]-S\left[ p(x|k)\right] \; .
\end{equation}
For short, we will call $\Delta_S(k)$ the {\em entropy change}. We
recover the Shannon information as the expected value of that
difference as
\begin{equation}
I_S= \langle \Delta_S(k) \rangle = \sum_k p(k) \Delta_S(k)\; 
\end{equation}
where Eve's knowledge $k$ occurs with probability $p(k)$.  If we are
able to give a bound on $\Delta_S(k)$ for a specific run of the
quantum key distribution experiment then this is  a stronger statement
than a bound a the Shannon information: we guarantee
not only security on average but make a statement on a specific key,
as required for secure communication.

The challenge for the theory of quantum cryptography is to provide a
statement like the following one: If one finds $e$ errors in a sifted
key of length $n_{\rm sif}$ then, after error correction under an
exchange of $N_{\rm rec}$ bits of redundant information, a new key of
length $n_{\rm fin}$ can be distilled on which, with probability
$1-\alpha$, a potential eavesdropper achieves an entropy change of
less than $\Delta_{\rm tol}$. Here $\Delta_{\rm tol}$ has to be chosen
in view of the application for which the secret key is used for.  It
is not necessary that each realization of a sifted key leads to a
secret key; the realization may be rejected with some probability
$\beta$. In that case Alice and Bob abort the attempt and start anew.

The final goal is to provide the security statement taking into
account the real experimental situation. For example, no real channel
exist which fulfill the axiom of faithfulness. There is the danger
that an eavesdropper can separate Alice and Bob and replace the public
channel by two channels: one from Alice to Eve and another one from
Eve to Bob. In this separate world scenario Eve could learn to know
the full key without causing errors.  She could establish different
keys with Alice and Bob and then transfer effectively the messages
from Alice to Bob. This problem can be overcome by {\it
authentication} \cite{wegman81a}.  This technique makes it possible
for a receiver of a message to verify that the message was indeed send
by the presumed sender. It requires that sender and receiver share
some secret knowledge beforehand. It should be noted that it is not
necessary to authenticate all individual messages sent along the
public channel. It is sufficient to authenticate some essential steps,
including the final key, as indicated below. In the presented
protocol, successful authentication verifies at the same time that no
errors remained after the key reconciliation.  The need to share a
secret key beforehand to accomplish authentication reduces this scheme
from a quantum key distribution system to a quantum key growing
system: from a short secret key we grow a longer secret key.  On the
other hand, since one needs to share a secret key beforehand anyway,
one can use part of it to control the flow of side-information to Eve
during the stage of key reconciliation in a new way. With
side-information we mean any classical information about the
reconciled key leaking to the eavesdropper during the reconciliation.

Another problem is that in a real application we can not effectively
create single photon states. Recent developments by Law and Kimble
\cite{law97a} promise such sources, but present day experiments use
dim coherent states, that is coherent pulses with an expected photon
number of typically $1/10$ per signal. The component of the signal
containing two or more photon states, however, poses problems.  It is
known that an eavesdropper can, by the use of a quantum non-demolition
measurement of the total photon number and splitting of signals, learn
with certainty all signals containing more than one photon without
causing any errors in the sifted key. If Eve can get hold of an ideal
quantum channel this will lead to the existence of a maximum value of
loss in the channel which can be tolerated \cite{yuen96a,huttner95a}.
It is not known at present whether this QND attack, possibly combined
with attacks on the remaining single photons, is the optimal attack
but it is certainly pretty strong.

The eavesdropper is restricted in her power to interfere with the
quantum signals only by quantum mechanics. In the most general
scenario, she can entangle the signals with a probe of arbitrary
dimensions, wait until all classical information is transmitted over
the public channel, and then make a measurement on the auxiliary
system to extract as much information as possible about the key. Many
papers, so far, deal only with single photon signals.  At present
there exists an important claim of a security proof in this scenario
by Mayers \cite{mayers98a}. However, the protocol proposed there is,
up to now, far less efficient than the here proposed one. Other
security proofs extend to a fairly wide class of eavesdropping
attacks, the coherent attacks \cite{biham98a}.

In this paper I will give a solution to a restricted problem. The
restriction consists of four points:
\begin{itemize}
\item The eavesdropper attacks each signal individually, no {\it
coherent or collective attacks} take place.
\item The signal states consist, indeed, of two pairs of orthogonal
single photon states so that two states drawn from different  pairs have
overlap probability $1/2$.
\item Bob uses detectors of identical detection efficiencies.
\item The initial key shared by Alice and Bob is secret, that is the
eavesdropper has negligible information about it. Using the part of
the key grown in a previous quantum key growing session is assumed to
be safe in this sense.
\end{itemize}
Within these assumptions I give a procedure that leads with some
a-priori probability $\beta$ to a key shared by Alice and Bob. If
successful, the key is secure in the sense that with probability
$(1-\alpha)$ any potential eavesdropper achieved an entropy change
less than $\Delta_{tol}$. In contrast to all other work on this
subject, this procedure takes into account that the eavesdropper does
not necessarily transmit single photons to the receiver; she might use
multi-photon signals to manipulate Bob's detectors. The procedure
presented here might not be optimal, but it is certifiable safe within
the four restrictions mentioned before.

 It should be pointed out that coherent eavesdropping attacks are at
 present beyond our experimental capability. Alice and Bob can
 increase the difficulty of the task of coherent or collective
 eavesdropping attacks by using random timing for their signals
 (although here one has to be weary about the error rate of the key)
 or by delaying their classical communication thereby forcing Eve to
 store her auxiliary probe system coherently for longer time. There is
 an important difference between the threat of growing computer power
 against classical encryption techniques and the growing power of
 experimental skills in the attack on quantum key distribution: while
 it is possible to decode today's message with tomorrow's computer in
 classical cryptography, you can not use tomorrow's experimental
 skills in eavesdropping on a photon sent and detected today. It is
 seems therefore perfectly legal to put some technological
 restrictions on the eavesdropper. This might be, for example, the
 restriction to attacks on individual system, or even the restriction
 to un-delayed measurements. For the use of dim coherent states one
 might be tempted to disallow Eve to use perfect quantum channels and
 to give her a minimum amount of damping of her quantum channel. The
 ultimate goal, however, should be to be {\it able} to cope without
 those restrictions.

The structure of the paper is as follows. In section \ref{howto} I
present the complete protocol on which the security analysis is
based. Then, in section \ref{elements} I discuss in more detail the
various elements contributing to the protocol. The heart of the
security analysis is presented in section \ref{expected} before I
summarize in section \ref{analysis} the efficiency and security of the
protocol.

\section{How to do quantum key growing}
\label{howto}
The protocol presented here is a suitable combination of the
Bennett-Brassard protocol, reconciliation techniques and
authentication methods. I make use of the fact that Alice and Bob have
to share some secret key beforehand. Instead of seeing that as a
draw-back, I make use of it to simplify the control of the
side-information flow during the classical data
exchange. Side-information might leak to Eve in the form of parity
bits, exchanged between Alice and Bob during reconciliation, or in the
form of knowledge that a specific bit was received correctly or
incorrectly by Bob. The side-information could be taken care of this
during the privacy amplification step using the results of
\cite{cachin97a}. Here I present for clarity a new method to avoid any
such side-information which correlates Eve's information about
different bits (as parity bits do which are typically used in
reconciliation) by using secret bits to encode some of the classical
communication.

The notation of the variables is guided by the idea that $n_x$ denotes
numbers of bits, especially key length at various stage, $N_x$ denotes
numbers of secure bits used in different steps of the protocol,
$\beta_i$ denote probabilities of failing to establish a shared key,
$\alpha_i$ denote failure probabilities critical to the safety of an
established key, while $\gamma$ denotes the probability that Alice and
Bob, unknown to themselves, do not even share a key.  Quantities
$\overline{x}$ or $\left< x \right>$ denote expected values of the
quantity $x$.

The protocol steps and their achievements are:
\begin{enumerate}
\item Alice sends a sufficient number of signals to Bob to generate a
sifted key of length $n_{\rm sif}$.
\item Bob notifies Alice in which time slot he received a signal.
\item Alice and Bob make a ``time stamp'' allowing them to make sure
that the previous step has been completed before they begin the next
step. This can be done, for example, by taking the time of
synchronized clocks after step $2$ and to include this time into the
authentication procedure.
\item Alice sends the bases used for the signals marked in the second
step to Bob.
\item Bob compares this information with his measurements and
announces to Alice the elements of the generalized sifted key of
length $n_{\rm sif}$. The generalized sifted key is formed by two groups
of signals. The first is the sifted key of the BB84 protocol formed by
all those signals which Bob can unambiguously interpret as a
deterministic measurement result of a single photon signal state. The
second group consists of those signals which are ambiguous as they can
not be thought of as triggered by single photon signals. If two of Bob
detectors (for example monitoring orthogonal modes) are triggered,
then this is an example of an ambiguous signal. The number of these
ambiguous signals is denoted by $n_D$.

The announcement of this step has to be included into the authentication.

\item Reconciliation: Alice sends, in total, $N_{\rm rec}$ encoded
 parity-check bits over the classical channel to Bob as a key
 reconciliation.  Bob uses these bits to correct or to discard the
 errors. During this step he will learn the actual number of errors
 $n_{\rm err}$.  The probability that an error remains in the sifted
 key is given by $\beta_1$.  Depending on the reconciliation scheme,
 Eve learns nothing in this step, or knows the position of the errors,
 or knows that Bob received all the remaining bits correctly.
\item From the observed number of errors $n_{\rm err}$ and of
 ambiguous non-vacuum results $n_D$ Bob can conclude, using a theorem
 by Hoeffding, that the expected disturbance measure
 $\overline{\epsilon} = \left< \frac{n_{\rm err} + w_D n_D}{n_{\rm
 sif}} \right> $ is, with probability $1-\alpha_1$, below a suitable
 chosen upper bound $\overline{\epsilon}_{\rm max}$. With probability
 $1- \beta_2$ they find a value for $\alpha_1$ which allows them
 to continue this protocol successfully. Here $w_D$ is a weight factor
 fixed later on.
\item Given the upper bound on the disturbance rate
 $\overline{\epsilon}_{\rm max}$, Alice and Bob shorten the key by a
 fraction $\tau$ during privacy amplification such that the Shannon
 information on that final key is below $I$. The shortening is
 accomplished using a hash function \cite{wegman81a} chosen at random.
 To make a statement about the entropy change $\Delta_S(k)$ Eve
 achieved for this particular transmission they observe that this
 change is with probability $1-\alpha_2$ less than $\Delta_{\rm
 tol}$. The probability $\alpha_2$ can be estimated by $\alpha_2
 <\frac{I}{ \Delta_{\rm tol}}$.
\item In the last step Alice chooses at random a suitable hash
 function which she transmits encrypted to Bob using $N_{\rm aut}/2$
 secret bits. Then she hashes with that function her new key, the time
 from step $3$, and the string of bases from step $5$ into a short
 sequence, called the {\it authentication tag}, The tag is sent to Bob
 who compares it with the hashed version of his key. If no error was
 left after the error correction the tags coincide.This step is
 repeated with the roles of Alice and Bob interchanged.  If Bob
 detects an error rate too high to allow to proceed with the protocol,
 he does not forward the correct authentication to Alice. The
 probability Eve could have guessed the secret bits used by Alice or
 by Bob to encode their hashed message is given by $\alpha_3$. The
 probability that a discrepancy between the two versions of the key
 remains undetected is denoted by $\gamma$.
\end{enumerate}

The {\it probability of detected failure} is $\beta$ with $\beta <
 \beta_1 + \beta_2$ and this failure does not compromise the
 security. In the case of success Alice and Bob can now say that, at worst, 
 with a {\it probability of undetected failure} (failure of security)
 of $\alpha$ (with $\alpha < \alpha_1 +\alpha_2 + \alpha_3$) the
 eavesdropper can achieve an entropy change for the final key which is
 bigger than $\Delta_{\rm tol}$. The remaining probability $\gamma$
 describes the probability that Alice and Bob do not detect that they
 do not even share a key.

Note that the final authentication is made symmetric so that no
exchange of information over the success of that step is
necessary. Otherwise a party not comparing the authentication tags
could regard the key as safe in a separate-world scenario. More
explanation about the authentication procedure can be found in section
\ref{authentication}.  The classical information becoming available to
Eve during the creation of the sifted key will be taken care of in the
calculations of section \ref{expected}.

The public channel is now used for the following tasks:
\begin{itemize}
\item creation of the sifted key, where Eve learns which signals
 reached Bob and from which signal set each signal was chosen from,
\item transmission of encrypted parity check bits, on which Eve learns
nothing,
\item for bi-directional reconciliation methods: feedback concerning
the success of parity bit comparisons (see following section),
\item for reconciliation methods which discard errors: the location of
bits discarded from the key,
\item announcement of the hash function chosen in this particular
realization,
\item transmission of the encrypted hash function for authentication
and of the unencrypted authentication tags.
\end{itemize}

The main subject of this paper is to give the fraction $\tau$ by which
the key has to be shortened to match the security target as a
function of the upper bound on the disturbance
$\overline{\epsilon}_{\rm max}$. The estimation has to take care of
all information available to Eve by a combination of measurements on
the quantum channel and classical information overheard on the public
channel. This classical information depends on the reconciliation
procedure used. The nature of this information might allow Eve to
separate the signals into subsets of signals, for example those being
formed by the signals which are correctly (incorrectly) received by
Bob, and to treat them differently.

 The knowledge of the specific hash function is of no use to Eve in
construction of her measurement on the signals. This is a result of
the assumption that Eve attacks each signal individually and that the
knowledge of the hash functions tells Eve only whether a specific bit
will count towards the parity bit of a signal subset or not. She only will
learn how important each individual bit is to her. If the bit is not
used then it is too late to change the interaction with that bit to
avoid unnecessary errors, since the damage by interaction has been
done long before.  If it is used, then Eve intends to get the best
possible knowledge about it anyway. This situation might be different
for scenarios which allow coherent attacks.

\section {Elements of the quantum key growing protocol}
\label{elements}
In this section I  explain in more detail the steps of the
quantum key growing protocol. Special attention is given to the
security failure probabilities $\alpha_i$, limiting the security
confidence of an established shared key, and to the failure
probabilities $\beta_i$, limiting the capability to establish a shared
key.

\subsection{Generation of the generalized sifted key}
Elements of the generalized sifted key are signals which either can be
unambiguously interpreted as being deterministicly detected, given the
knowledge of the polarization basis, or which trigger more than one
detector.  We think of detection set-ups where detectors monitor one
relevant mode each.  Due to loss it is possible to find no photon in
any mode. Since Eve might use multi-photon signals we may find photons
in different monitored modes simultaneously, leading to ambiguous
signals since more than one detector gives a click. Detection of
several photons in {\it one} mode, however, is deemed to be an
unambiguous result. (See further discussion in section
\ref{evesinteraction}.) In practice we will not be able to distinguish
between one or several photons triggering the detector. The length of
the sifted key accumulated in that way is kept fix to be of length
$n_{\rm sif}$.

\subsection{Reconciliation}
For the reconciliation we have to distinguish two main classes of
procedures: one class corrects the errors using redundant information
and the other class discards errors by locating error-free subsections
of the sifted key.  The class of error-correcting reconciliation can
be divided in two further subclasses: one subclass uses only
uni-directional information flow from Alice to Bob while the second
subclass uses an interactive protocol with bi-directional information
flow.

The difference between the three approaches with respect to our
protocol shows up in the number of secret bits they need to reconcile
the string, the length of the reconciled string, and the probability
of success of reconciliation. For experimental realization one should
think as well of the practical implementation. For example,
interactive protocols are very efficient to implement
\cite{brassard93a}. To illustrate the difference I give examples for
the error correction protocols.

The benchmark for efficiency of error correction is the Shannon
limit. It gives the minimum number of bits which have to be revealed
about the correct version of a key to reconcile a version which is
subjected to an error rate $e$. This limit is achieved for 
large keys and the error correction probability approaches then
unity. The Shannon limit is given in terms of the amount of Shannon
information $I_S(e)$ contained in the version of the key affected by
the error rate $e$. For a binary channel, as relevant in our case,
this is given by
\begin{equation}
\label{binaryshannon}
I_S(e) = 1+ e \log e + (1-e) \log (1-e) \; .
\end{equation}
The minimum number of bits needed, on average, to correct a key of
length $n$ affected by the error rate $e$ is then given by
\begin{equation}
n_{\rm min} = n \; \left\{1-I_S(e)\right\} \; .
\end{equation}
As mentioned before, perfect error correction is achievable only for 
$n \to \infty$.

\subsubsection{Linear Codes for error correction.}
Linear codes are a well-established technique which can be viewed in a
standard-approach as attaching to each $k$-bit signal a number of
$(n-k)$ bits of linearly independent parity-check bits making it in
total a $n$-bit signal.  The receiver gets a noisy version of this
n-bit signal and can now in a well-defined procedure find the
most-likely $k$-bit signal. Linear codes which will safely return the
correct $k$-bit signal if up to $f$ of the $n$ bits were flipped by
the noisy channel are denoted by $[n,k,d]$ codes (with $d=2f+1$). If the
signal is affected by more errors then these will be corrected with
less than unit probability.

This technique can be used for error correction. Alice and Bob
partition their sifted key into blocks of size $k$. For each block
Alice computes the extra $n-k$ parity bits, encodes them with secret
bits and sends them via the classical channel to Bob. Bob then
corrects his block according to the standard error correction
technique.  This procedure could be improved, since the $[n,k,d]$ codes
are designed to cope with the situation that even the parity bits
might be affected by noise. One can partly take advantage of the
situation that these bits are transmitted correctly. However,
non-optimal performance is not a security hazard.

The search for an optimal linear code is beyond the scope of this
paper. To illustrate the problem I present as specific example
the code $[512,422,21]$. It uses $90$ redundant parity bits to
protect a block of $422$ bits against $10$ errors. So how does this
linear code perform if we use it to reconcile a string of $n_{\rm
sif}=10128$ bits which are affected by an error rate of $1 \%$?  It
can be shown that this string will be reconciled with a probability of
$(1-\beta_1) =0.908$ at an expense of $N_{\rm rec}=2160$ secret
bits. The practical implementation of a code as long as this one is,
however, rather problematic from the point of view of computational
resources.
In comparison, in the Shannon limit we need to use $819$
bits for this task.

\subsubsection{Interactive error correction}
An interactive error correction code was presented by Brassard and
Salvail in \cite{brassard93a}. This code is reported to correct a key
with an error rate of $1\%$ and length $n_{\rm sif}=10000$ at an average
expense of $N_{\rm rec} = 933$ bits. No numbers for $\beta_1$ are given,
but in several tries no remaining error was found. This protocol
operates acceptable close to the Shannon limit which tells us that we
need at least $808$ bits to correct the key.

\subsubsection{Situation after reconciliation}
After reconciliation Alice and Bob share with probability
$(1-\beta_1)$ the same key. The eavesdropper gathered some
information from measurements on the quantum channel. The information
she gained from listening to the public channel puts her now into
different positions depending on the reconciliation protocol. In case
errors are discarded, she knows that all remaining bits in the
reconciled string were received correctly by Bob during the quantum
transmission. If an uni-directional error correction protocol is
used, then listening to the public channel during reconciliation
does not give Eve any extra hints. The interactive error correction
protocol, however, leaks some information to Eve about
the position of bits which  were received incorrectly
by Bob during the quantum protocol. We will have to take this into
account later on. There we take the view that Eve knows the
positions of all errors exactly.

A difference between correcting and discarding errors is that,
naturally, discarding errors will lead to a shorter reconciled string
of length $n_{\rm rec} < n_{\rm sif}$, while the length of the key
does not change during error correction so that $n_{\rm rec} = n_{\rm
sif}$. Common to all schemes is that Alice and Bob know the precise
number of errors which occurred (provided the reconciliation
worked). When they discard parts of the sifted key they can open up
the discarded bits and learn thereby the actual number of errors
(although in this case an additional problem of authentication
arises), and when they correct errors Bob knows the number of
bit-flips he performed during error correction. This is just the
number of errors of the sifted key.

Contrary to common belief it is therefore not necessary to sacrifice
elements of the sifted key by public comparison to determine or
estimate the number of occurred errors.

\subsection{Privacy amplification and the Shannon information on final key}
In previous work it has been shown that for typical error rates in an
experimental set-up the eavesdropper could gain, on average,
non-negligible amount of Shannon information on the reconciled key
\cite{fuchs97a,nl96a}. This means that we can not use it as a secret
key right away. Classical coding theory shows a way to distill a final
secret key from the reconciled key by the method of privacy
amplification \cite{bennett95a}. As a practical implementation of the
hashing involved, the secret key is obtained by taking $n_{\rm fin} $
parity bits of randomly chosen subsets of the $n_{\rm rec}$ bits of
the reconciled string.  The choice of the random subsets is made only
at that instance and changes for each repetition of the key growing
protocol. This shortening of the key to enhance the security of the
final key is common to all other approaches that deal with the
security of quantum cryptography, for example by Mayers
\cite{mayers98a} or Biham et al \cite{biham98a}. However, it differs
the way to determine the fraction $\tau$ by which the key has to be
shortened. In the case of individual eavesdropping attacks we can go
via the collision probability as described below
\cite{bennett95a}. When we consider joint or collective attacks it is
not possible to take this approach due to correlation between the
signals which possibly allows Eve to gain an advantage by delaying her
measurement until she learns to know the specific parity bits entering
the final key.

In the first step we give the main formulas of privacy amplification
and introduce the parameter $\tau_1(\overline{\epsilon})$. This
parameter indicates the fraction by which the key has to be shortened
such that the {\it expected} eavesdropping information on the final
key is less than 1 bit of Shannon information.  It is given as a
function of Eve's acquired {\it collision probability}. Any additional
bit by which the key is shortened leads to an exponential decrease of
that expected Shannon information.

We denote by $z$ the final key of length $n_{\rm fin}$, by $x$ the
reconciled key of length $n_{\rm rec}$ and by $y$ the accumulated
knowledge of the eavesdropper due to her interaction with the signals
and the overheard classical communication via the public channel. We
keep separately the {\it hash function} $g$ which, for example,
describes the subsets whose parity bits form the final key. This hash
function is part of Eve's knowledge in each realization. Eve's
knowledge is expressed in a probability distribution $p(z|g,y)$, that
is the probability that $z$ is the key given Eve's measurement results
and side-information on the key. In a trivial extension of the
starting equation of \cite{bennett95a} we find that the Shannon
information $\tilde{I}$, averaged over the hash functions, is bounded
by
\begin{equation}
I \equiv \langle \tilde{I} \rangle_{g} \leq n_{\rm fin} + \log
\langle p_c^z(g,y) \rangle_{y,g}
\end{equation}
with the collision probability on the final key defined as $p_c^z(g,y)
= \sum_z p^2(z|g,y)$. The collision probability $\langle p_c^z(g,y)
\rangle_g$ on the final key, averaged with respect to $g$, is bounded
by the collision probability $p_c^x(y) = \sum_x p^2(x|y)$ on the
reconciled key as
\begin{equation}
\langle p_c^z(g,y) \rangle_g < 2^{-n_{\rm fin}} \left( 2^{n_{\rm fin}}
p_c^x(y) + 1\right) \; .
\end{equation}
This can be trivially extended to an inequality for $\langle
p_c^z(g,y) \rangle_{y,g}$ resulting in
\begin{equation}
\langle p_c(g,y) \rangle_{g,y} < 2^{-n_{\rm fin}} \left( 2^{n_{\rm
fin}} \langle p_c^x(y) \rangle_{y} + 1\right) \; .
\end{equation}
This allows us to give the estimate
\begin{equation}
\label{ipclink}
I \leq \log \left( 2^{n_{\rm fin}} \langle p_c^x(y) \rangle_{y}
+ 1 \right)\;
\end{equation}
bounding the eavesdropper's expected Shannon information by her
expected collision probability on the sifted key and the length of the
final key.

We can reformulate the estimate (\ref{ipclink}) by introducing the
fraction $\tau_1$. If we shorten the reconciled key by this fraction
then Eve's expected Shannon information is just one bit on the whole
final key. Therefore we find
\begin{equation}
\label{taublub}
\tau_1=1 + \frac{1}{n_{\rm rec}} \log \langle p_c^x(y) \rangle_{y} \; .
\end{equation}
We introduce the security parameter $n_S$ as the number of bits by which the
final key is shorter than prescribed by the fraction
$\tau_1$. This security parameter $n_S$ is implicitly defined by
\begin{equation}
n_{\rm fin} = (1-\tau_1)\; n_{\rm rec} - n_S \; .
\end{equation}
With the definitions of $\tau_1$ and $n_S$ we then find \cite{bennett95a}
\begin{equation}
\label{iestimate}
I \leq \log ( 2^{-n_S} +1) \approx \frac{2^{-n_S}}{\ln 2} \; .
\end{equation}
From this relation we see that the total amount of Eve's expected
Shannon information on the final key decreases exponentially with the
security parameter $n_S$.  The main part of this paper will be to
estimate $\langle p_c^x(y) \rangle_{y}$ for various scenarios as a
function of the expected disturbance rate $\overline{\epsilon}$ to
estimate $\tau_1$ and with that to estimate $I$ as a function
$\overline{\epsilon}$.

\subsection{From expected quantities to specific quantities}
In the previous section we showed that once we know the expected
disturbance rate $\overline{\epsilon}$ and the functional dependence
of $\tau_1(\overline{\epsilon})$, we can estimate the eavesdropper's
Shannon information $I$ on the final key in dependence  of $n_S$
via equation (\ref{iestimate}). In this section we now show how to
link the observed error rate to the expected error rate and how to
estimate the entropy change $\Delta_S$ in a single run from the
expected Shannon information $I$.

\subsubsection{From the measured error rate to the expected error
rate} Alice and Bob establish a generalized sifted key of length
$n_{\rm sif}$. During reconciliation of the sifted key Bob learns 
the actual number of errors $n_{\rm err}$ of unambiguous signals while
he already knows the number $n_D$ of ambiguous signals.  Our
definition of disturbance is here 
\begin{equation}
\epsilon =\frac{n_{\rm err} + w_D n_D}{n_{\rm rec}}
\end{equation}
 with $w_D$ as adjustable weight parameter for ambiguous signals to be
chosen in a suitable way.  We will present in section
\ref{multisection} a model for which we can choose $w_D = 1/2$. In the
case of error correction we have to correct even the ambiguous signals
to keep the number $n_{\rm sif}$ fixed and to keep control about the
disturbance. The reason is we need to formulate a measure of
disturbance per element of the reconciliated key which is
bounded. This is possible for correction of errors. In the case of
discarding errors the number of errors and ambiguous results per
remaining bit is unbounded and we fail to be able to give a bound on
$\overline{\epsilon}$ from the measured values.

Therefore we restrict ourselves to the case of corrected errors where
we find the length $n_{\rm rec}$ of the reconciled string to be equal
to the length $n_{\rm sif}$ of the generalized sifted key. In this
situation the measured disturbance is given by $\epsilon_{\rm meas}
=\frac{n_{\rm err} + w_D n_D}{n_{\rm sif}}$. Since $n_{\rm sif}$ is
kept fixed the expected disturbance is given by $ \overline{\epsilon}
= \frac{\left< n_{\rm err} + w_D n_D\right>}{n_{\rm sif}}$.  From the
measured value $\epsilon_{\rm meas}$ we estimate the average
disturbance parameter $\overline{\epsilon}$.

To make the role of $\overline{\epsilon}$ clear it should be pointed
out that any given eavesdropping strategy will lead to an expected
error probability $\overline{\epsilon}$ while the actually caused and
observed error rate can be much lower for an individual run of the
protocol. For example, think of an intercept/resend protocol as in
\cite{huttner95a} where Eve has her lucky day and measures, by chance,
all signals in the appropriate bases. This is not very likely, but the
treatment presented here takes care of this possibility.

 In an application of a theorem by Hoeffding \cite{hoeffding63a},
which has been used already in \cite{biham98a}, we find an estimate of
the number $\left< n_{\rm err} + w_D n_D\right>$ from the actually
measured number $ n_{\rm err}+ w_D n_D$ for a total number of $n_{\rm
sif}$ signals as
\begin{equation}
\label{expest}
\label{alpha1}
\left< n_{\rm err}+ w_D n_D\right> < n_{\rm err}+ w_D n_D + n_{\rm
sif} \delta
\end{equation}
with probability
\begin{equation}
\label{succprob}
 (1- \alpha_1) 
 > 1-\exp(-2 n_{\rm sif} \delta^2) 
\end{equation}
as long as $w_D \leq 1$. For $w_D \geq 1$ we have to replace equation
(\ref{succprob}) by $(1- \alpha_1) > 1-\exp(-\frac{2 n_{\rm sif}
\delta^2}{w_D^2})$.  This means that we can give a bound on the
expected disturbance parameter $\overline{\epsilon}$ from the observed
quantities $n_D$ and $n_{\rm err}$ within a certain confidence limit.
To give a numeric example we choose $w_D = 1/2$ (see section
\ref{multisection}) and refer to the situation reported by Marand and
Townsend \cite{marand95a}. There an experiment is presented which can
create a sifted key of length $n_{\rm sif} = 1.4 \times 10^{-3} n$
from an exchange of $n$ quantum signals at an error rate of $1.2 \% $
with a negligible amount of ambiguous signals. Then the choice of
$\delta = 0.038$ and a sampling with $n= 10^7$ leads to a reconciled
key of length $n_{\rm sif} = 1.4 \times 10^4$ with a value of
$\alpha_1 \approx 10^{-18}$. This is the probability that the expected
disturbance parameter $\overline{\epsilon}$ in a typical realization
of the key transfer is less than a maximal value of
$\overline{\epsilon}_{\rm max}=0.05$. The value
$\overline{\epsilon}_{\rm max}$ will be used in privacy amplification.
An eye has to be kept on the sampling time. With the experiment
described in \cite{marand95a} it will take about $10$ seconds to
establish the sifted key. An example for smaller samples is the choice
of $n = 10^{5}$ and $\delta=0.4$ which leads for the same system to a
reconciled key of length $n_{\rm sif} = 140$ and $\alpha_1 \approx
10^{-19}$, $\overline{\epsilon}_{\rm max}=0.412$. The probability
$\beta_2$ to fail to achieve a satisfactory level of confidence at
this stage is in most cases negligible in comparison to the failure of
reconciliation. It should be noted that these numbers give a rough
guidance only, since the experiment does not use single-photon
signals.

\subsubsection{Expected information and information in specific realization}
 We still need to link the change of Shannon entropy $\Delta_S$ on the
final key in an {\it individual} realization of the protocol with a given
probability to the Shannon information $I$, that is over the average
over many realizations.  The key is thought of as unsafe if the
eavesdropper achieves an entropy change bigger than $\Delta_{\rm tol}$
in a specific realization. This happens at most with probability
$\alpha_2$ which is bounded implicitly by $I > \alpha_2 \Delta_{\rm
tol}$ leading to
\begin{equation}
\label{alpha2}
\alpha_2 < \frac{I}{\Delta_{\rm tol}}= \frac{\log (2^{-n_S} +
1)}{\Delta_{\rm tol}} \approx \frac{2^{-n_S}}{\Delta_{\rm tol} \ln 2}
\end{equation}
So the knowledge of an estimate for $I$ and the prescription of an
acceptable value of $\Delta_{\rm tol}$ gives us the probability
$1-\alpha_2$ of secrecy of the key.

\subsection{Authentication}
\label{authentication}
The tools of the previous sections allow Alice and Bob to construct a
common secret key provided that their classical channel is
faithful. Since channels with that property, as such, do not exist, we
need to authenticate the procedure to make sure that Alice and Bob
actually {\it share} the new key.  Authentication can protect at the
same time against errors which survived the reconciliation step and
against an eavesdropping attack with a ``separate world'' approach.

It is essential to make sure that Eve has no influence on the choice
of bits entering the generalized sifted key exceeding the power to
manipulate the quantum channel.  The time-stamp step $3$ in the
protocol assures us that there is no point in Eve faking the public
discussion up to that point since she gained no additional information
about the signals so far, especially no information about the
polarization basis.

The following sequence of bases for the successful received signals
sent from Alice and Bob does not need to be authenticated as well
since Eve can not bar corresponding signals from the sifted key
without knowing Bob's measurements as well. However, the message
describing which bits finally form the generalized sifted key needs to
be authenticated since Eve is now in the position to bar signals from
the sifted key she shares with Alice by manipulation of the contents
of the message \cite{dusek}.

The subsequent reconciliation protocol need not to be authenticated if
we authenticate the final key. The reason for that is that the
previous steps fixed the reconciled key as the generalized sifted key
in Alice's version. If Eve tampers with the reconciliation protocol
then Bob will fail correct his key so that it becomes equal to Alice's
key. Authentication of the final key will therefore be sufficient to
protect against tampering with the public channel in this step. It
doubles at the same time to protect against incomplete reconciliation.

To summarize, we need to authenticate the string identifying the
elements of the sifted key within the received signals, the time
stamp, and the final key. The length of this string is roughly $m
\approx 2 n_{\rm sif}$. The authentication is done in the following
way which is based on the authentication procedure of Wegman and
Carter \cite{wegman81a}:

 Alice chooses a hash-function of approximate length $N_{\rm aut}/2 =
 4 t \log m$ and sends it encrypted to Bob. Both evaluate the hashed
 version of the message, the tag, of length $t$. Alice sends the tag
 via the public channel to Bob. If the tags coincide then this step is
 repeated with the role of Alice and Bob interchanged. With this
 symmetric scheme we make sure that neither Alice nor Bob can be
 coaxed into a position where they think that authentication succeeded
 when it in fact failed.  The probability that Eve could fake the
 authentication is given by
\begin{equation}
\label{alpha3}
\alpha_3 =2^{-t+1} \; .
\end{equation}
 This is at the same time the probability that two distinct final keys
lead to the same hashed key. Any remaining error in the final key will
therefore lead with probability $1-\alpha_3$ to a failure of the
authentication.

\section{Expected collision probability and expected error rate}
\label{expected}
This section represents  the major input of physics to the
quantum key growing protocol. The aim is to put an upper bound on the
expected average collision probability Eve obtains on the
reconciliated key as a function of an average disturbance rate her
eavesdropping strategy inflicted on the signals. This is done for two
methods of reconciliation, correcting or deleting errors. The result
will allow us to give values for the parameter $\tau_1(\overline{\epsilon})$.

\subsection{Collision probability on individual signal}
The  collision probability on the reconciled key is defined by
\begin{equation}
p_c^x(y)= \sum_x p^2(x|y) \; .
\end{equation}
We assume that the signal sent by Alice are statistically independent
of each other and Eve interacts with and performs measurements on each
bit individually. Furthermore, we avoid side-information which
correlates signals by the use of secret bits in the reconciliation
step.  Therefore the conditional probability function $p(x|y)$ for $x$
being the key given Eve's knowledge $y$ factorises into a product of
probabilities for each signal. With that the expected collision
probability factorises as well into a product of the expected
collision probability for each bit. We denote by $p_c^x$ the expected
collision probability on one bit so that
\[
\langle p_c^x(g,y) \rangle_y = \left(p_c^x \right)^{n_{\rm rec}}
\]
Furthermore, we denote by the index $\alpha \in
\{+,\times\}$ the two conjugate bases (e.g.~horizontal or
vertical polarization for single photons) used to encode the signals,
 by $\Psi \in \{0,1\}$ the logical values, and by $k$  the possible
outcomes of Eve's measurement. This leads to an
expression of the expected collision probability, at this stage, as
\begin{equation}
 p_c^x  = \sum_{k, \Psi, \alpha} \frac{
p^2(\Psi_\alpha, k_\alpha)}{p(k_\alpha)} \; .
\end{equation}
We find for the parameter $\tau_1$ describing the shortening of
the key during privacy amplification from eqn.~(\ref{taublub})
\begin{equation}
\tau_1 = \log (2 p_c^x) \; .
\end{equation}

\subsection{Eve's interaction and detection description}
\label{evesinteraction}
The action of the eavesdropper can be described by a completely
positive map \cite{davies76a,kraus83a} acting on the signal density
matrices $\rho$ as
\begin{equation}
\label{cpmapping}
\tilde{\rho} = \sum_k A_k \rho A_k^\dagger
\end{equation}
where we can associate this interaction with a measurement by Eve of a
 {\it Probability Operator Measure} (POM) formed by the operators $F_k =
 A_k^\dagger A_k$. The operators $A_k$ are arbitrary operators mapping
 the Hilbert space of the signals to an arbitrary Hilbert space. The
 only restriction is that $\sum_k A_k^\dagger A_k $ gives the identity
 operator of the signal Hilbert space. The probability for occurrence
 of outcome $k$ is then given by $p(k) = {\rm Tr}(\rho F_k)$.  The action of
 Bob's detectors can be described by a POM on the resulting Hilbert
 space after Eve's interaction. Since the detection POM elements and
 the signal density operators can be represented by real matrices, we
 can assume the operators $A_k$ to be represented by real matrices as
 well. 

 This does not limit the generality of the approach, since the outcome
 corresponding to an operator $A_k = A_k^{\rm re} + i A_k^{\rm im}$,
 with real operators $A_k^{\rm re}$ and $A_k^{\rm im}$, is triggered
 with probability ${\rm Tr}(\rho { A_k^{\rm re}}^\dagger A_k^{\rm re})
 + {\rm Tr}(\rho { A_k^{\rm im}}^\dagger A_k^{\rm im})$ and the
 outcome probabilities for Bob's detection, corresponding to POM
 element $F$ if outcome $k$ of Eve's measurement is being triggered,
 is given by ${\rm Tr}(A_k^{\rm re}\rho { A_k^{\rm re}}^\dagger F) +
 {\rm Tr}(A_k^{\rm re}\rho { A_k^{\rm re}}^\dagger F)$. Since no
 cross-terms mixing $A_k^{\rm re}$ and $A_k^{\rm im}$ occur this means
 that using the two real operators $A_k^{\rm re}$ and $A_k^{\rm im}$,
 instead of $A_k = A_k^{\rm re} + i A_k^{\rm im}$, will not change the
 outcome probabilities of Bob's detectors but refines Eve's
 measurement.

\begin{figure}
\centerline{\psfig{width=7cm,file=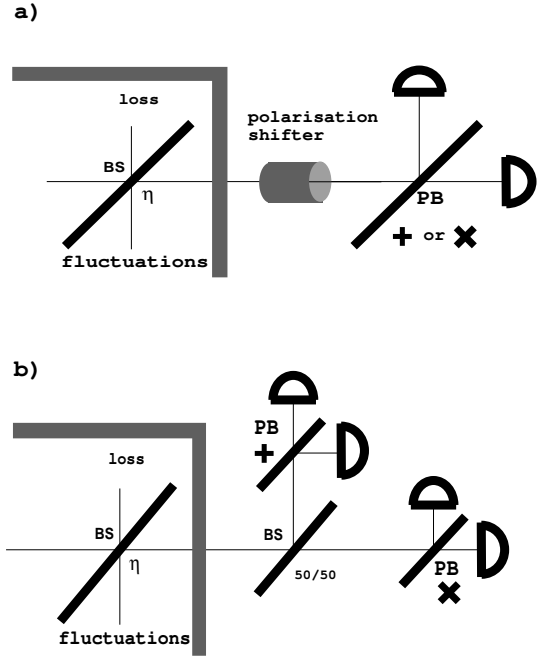}}
\caption{\label{detector} \it {\bf (a) Active device:} Bob's two
 detectors consist each of a polarizing beam splitter and an ideal
 detector. The polarizing beam splitter discriminates the two
 orthogonal linear polarized modes. Using a polarization shifter the
 polarization basis can be changed as desired. Detector efficiencies
 are modeled by a beam splitter which represents the loss and which is
 thought of as being part of the eavesdropper's strategy. This beam
 splitter can be seen as part of the quantum channel. {\bf (b) Passive
 device:} Here one uses two detection modules as presented in (a), one
 for each polarization basis. The central beam-splitter takes the task
 to ``switch'' between the two polarization analyzers.}
\end{figure}

Two typical detection set-ups are shown in figure \ref{detector}.  The
 active version consists of a polarization analyzer (two detectors
 monitoring each an output of a polarizing beam-splitter) and a phase
 shifter which effectively changes the polarization basis of the
 subsequent measurement.  Here one has actively to choose
 the polarization basis of the measurement. The passive device uses
 two polarization analyzers, one for each basis, and uses a
 beam-splitter to split the incoming signal the two polarization
 analyzers are used with equal probability for  detection.

One can represent the detectors by beam-splitters combined with
ideal detectors \cite{yurke85a}. Then the beam-splitters can be
thought to be responsible for the finite efficiency. Since all
detectors are assumed to be equal, the losses of all detectors
involved can be attributed to a single loss beam-splitter, which is
then thought of as being part of the transmission channel rather than
being part of the detection unit.

 We can use the idea of ideal detectors which measure each a POM with
 two elements, the projection operator onto the vacuum (no ``click'')
 and the projection on the Fock-subspaces with at least one photon
 (``click'').  The POM of the active and the passive set-up then
 contains the elements $F_{\rm vac}$, $ F_{0_+}$, $ F_{1_+}$, $
 F_{0_\times}$, $ F_{1_\times}$, $ F_D$. These are projections onto
 the vacuum, $F_{\rm vac}$, onto states with at least one photon in one
 of the four signal polarizations and none in the others, therefore
 leading to an unambiguous result, $F_{\Psi_\alpha}$, and onto the
 rest of the Hilbert space, that is onto all states containing at
 least one photon in the signal polarization and at least one in an
 orthogonal mode $F_D$. The first POM outcome manifests itself in no
 detector click at all, the following four give precisely one detector
 click, and the last one gives rise to at least two detectors being
 triggered. If we denote by $|n,m\rangle_\alpha$ the state which has
 $n$ photons in one mode and $m$ photons in the orthogonal
 polarization mode with respect to the polarization basis $\alpha$,
  use the abbreviation $E^{(0)}$ for the projector onto the vacuum and
 $E^{(n)}_{\Psi_\alpha}$ for the projector onto the state with $n$
 photons in the polarization mode corresponding to $\Psi_\alpha$, then
 the POM of detection unit (a) is given by
\begin{eqnarray}
F_{\rm vac} & = & E^{(0)} \\
 F_{\Psi_\alpha} &=& \frac{1}{2}
\sum_{n=1}^\infty E^{(n)}_{\Psi_\alpha} \nonumber \\
 F_D & = &
\frac{1}{2} \sum_{ n,m=1}^{\infty } |n,m \rangle_+
\langle n,m| + \frac{1}{2} \sum_{n,m=1}^{\infty }
|n,m \rangle_\times \langle n,m| \; . \nonumber
\end{eqnarray}

On the other hand, the passive detection scheme (b) is more
susceptible to signals containing more than one photon. It is
described by the POM
\begin{eqnarray}
\label{passivpom}
F_{\rm vac} & = & E^{(0)} \\
 F_{\Psi_\alpha} &=& \sum_{n=1}^\infty
\left(\frac{1}{2}\right)^n E^{(n)}_{\Psi_\alpha} \nonumber \\
 F_D & =
&\sum_{n=1}^\infty\left[\left(\frac{1}{2}- \left(\frac{1}{2}\right)^n
\right)\sum_{\Psi,\alpha} E^{(n)}_{\Psi,\alpha}\right] \nonumber \\
 & &  + \frac{1}{2}
\sum_{n,m=1}^{\infty } |n,m \rangle_+ \langle n,m|
+ \frac{1}{2} \sum_{n,m=1}^{\infty } |n,m
\rangle_\times \langle n,m| \; . \nonumber
\end{eqnarray}

The next idea concerns all detection set-ups where all elements of the
POM commute with the projections $E_n$ onto the subspaces of
total photon number $n$. In that case we find that Bob's measurement
on the final signal gives outcome $i \in \{ {\rm vac}, \Psi_\alpha, D \}$
with probability
$$
P_{\rm Bob}(i) := {\rm Tr}(\sum_k A_k \rho A_k^\dagger F_i) = {\rm
Tr}(\sum_{k,n} E_n A_k \rho A_k^\dagger E_n F_i)
$$
We can now replace the set of $A_k$'s by the set $A_{n,k}:=E_n A_k$
which still describes Eve measurement but for which each element maps
the Hilbert space of the signals to a Hilbert space with a fixed
photon number. Eve will now associate a POM element of her measurement
with each such $A_{n,k}$ thereby refining her POM and leading to an
increase of her knowledge. For short we write again $A_k$ for this
set, for which now the property is assumed that the signal arriving at
Bob's detection unit is an eigenstate of the total photon number
operator. We can divide the index set $K$ of $k$ into subsets
$K^{(n)}$ so that for each $ k \in K^{(n)}$ the operator $A_k$ maps
the one-photon Hilbert space of the signal into the $n$-photon
space. This is useful to distinguish contributions of signals with
different photon number.

We still have to discuss how to represent a delayed measurement in
this picture. A delayed measurement is performed in the way that Eve
brings an auxiliary system into contact with the signal so that they
evolve together under a controlled unitary evolution. Then the signal
is measured by Bob while Eve delays the measurement of her auxiliary
system until she has received all classical information exchanged over
the public channel. Having this knowledge, she picks  the optimal
measurement to be performed on her auxiliary system.  Classical
information useful to Eve is information that allows her to divide the
signals into subsets which should experience different treatment.  In
our situation this information is represented by the polarization
basis of the signal and, for bi-directional error correction, by the
knowledge whether the signal was received correctly by Bob.  We have
therefore to assume, for example, that Eve's delayed measurement is
characterized by the set of operators $A_k$ with $k \in K$, giving
rise to Eve's POM $F_k = A_k A_k^\dagger$, and which are applied to
the signals from the set $\alpha = \mbox{``$+$''}$ and a second set
$B_{k'}$ with $k' \in K'$, resulting in the POM $F'_{k'} = B_{k'}
B_{k'}^\dagger$, which are applied to the signals from the set $\alpha
= \mbox{``$\times$''}$. Of course, these two sets of operators can not
be chosen arbitrarily. The complete positive map has to be identical
for all density matrices $\rho$, that is
\begin{equation}
\tilde{\rho} = \sum_{k \in K} A_k \rho A_k^\dagger = \sum_{k' \in K'}
B_{k'} \rho B_{k'}^\dagger \; .
\end{equation}
Moreover, this equality  holds even for non-Hermitian matrices
$\rho$. We can combine this result with the partition into  $n$-photon
subspaces. Then we find that even the stronger statement
\begin{equation}
\label{ABequality}
 \sum_{k \in K^{(n)}} A_k \rho A_k^\dagger = \sum_{k' \in K'^{(n)}}
 B_{k'} \rho B_{k'}^\dagger \; .
\end{equation}
holds.  Before we go on to the derivation of the relation between
average disturbance and average collision probability I would like to
point out that this treatment takes into account the rich structure of
modes supported by optical fibers and the fact that detectors monitor
a multitude of modes.  As long as the detection POM commutes with the
projector onto the actually used signal mode, which is usually the
case, we can separate the action of the $A_k$ with respect to the
photon number in a similar way.

\subsection{Separation into $n$-photon contributions}
In this section we are going to present the disturbance measure
$\epsilon$ and the collision probability $\ p_c^x $ as sums over
contributions with different definite photon number $n$ arriving at
Bob's detector unit. We start from the definition of the disturbance
$\epsilon$. To allow some comparison between correcting and discarding
errors, we present a unified definition which defines, even for
discarded errors, a disturbance measure per bit of the reconciled
key. This definition is given by
\begin{equation}
\epsilon = \frac{n_{\rm err} + w_D n_D}{n_{\rm rec}} \; .
\end{equation}
Here $n_{\rm err}$ is the number of errors in the sifted key, $n_D$ is
the number of ambiguous results occurring and $n_{\rm rec}$ is the
number of bits in the reconciled string. The weight parameter $w_D$
for ambiguous signals will be fixed later on.  If we keep the size of
the reconciled key fixed, then the expectation value of $\epsilon$ is
described by
\begin{equation}
\overline{\epsilon} = \frac{p_{\rm err} + w_D p_D}{p_{\rm rec}} \;
\end{equation}
where $p_{\rm err}, p_D, p_{\rm rec}$ are the absolute probabilities
that a signal will, respectively, enter the sifted key as error, cause
an ambiguous result, or become an element of the reconciled key.  As
mentioned before, it should be noted, that no estimate
$\overline{\epsilon}$ from measured data can be easily presented in
the case of discarded errors.  We separate the contributions from the
different photon number signals as
\begin{equation}
\overline{\epsilon} = \sum_n \frac{p_{\rm rec}^{(n)}}{p_{\rm
 rec}}\frac{p_{\rm err}^{(n)} + w_D p_D^{(n)}}{p_{\rm rec}^{(n)}} =
 \sum_n \frac{p_{\rm rec}^{(n)}}{p_{\rm rec}}
 \overline{\epsilon}^{(n)} \; .
\end{equation}
where we have implicitly defined
\begin{equation}
\label{eps1def}
\overline{\epsilon}^{(n)} = \frac{p_{\rm err}^{(n)} + w_D
p_D^{(n)}}{p_{\rm rec}^{(n)}}
\end{equation}
as the n-photon contribution towards the disturbance measure. Now
$p_{X}^{(n)}$ are the conditional probabilities that a signal has
property $X$ while being transfered as $n$-photon signal between Eve
and Bob.  The total disturbance is given as sum over the n-photon
contribution weighted by the relative probability that a signal
arriving as an n-photon signal at Bob's detector will enter the
reconciled key.

If we discard errors, then we find for the relevant probabilities
(with $\overline{\Psi}$ as the complement to binary value $\Psi$)
\begin{eqnarray}
p_{\rm err}^{(n)} & = & \frac{1}{4} \sum_{k \in K^{(n)} \atop
\Psi,\alpha, } {\rm Tr}\left( A_k \rho_{\Psi_\alpha} A_k^\dagger
F_{\overline{\Psi}_\alpha}^{(n)} \right) \\ \label{refref} p_{\rm
rec}^{(n)} & = & \frac{1}{4} \sum_{k \in K^{(n)} \atop \Psi,\alpha, }
{\rm Tr}\left( A_k \rho_{\Psi_\alpha} A_k^\dagger F_{\Psi_\alpha}^{(n)}
\right) \\ p_D^{(n)} & = & \frac{1}{4} \sum_{k \in K^{(n)} \atop
\Psi,\alpha, } {\rm Tr}\left( A_k \rho_{\Psi_\alpha} A_k^\dagger F_D^{(n)}
\right) \; .
\end{eqnarray}
If we correct errors, then the probability for a signal to enter the
reconciled key differs from equation (\ref{refref}) and is, instead,
given by
\begin{eqnarray}
p_{\rm rec}^{(n)} & = & \frac{1}{4} \sum_{k \in K^{(n)} \atop
\Psi,\alpha, \Psi' } {\rm Tr}\left( A_k \rho_{\Psi'_\alpha} A_k^\dagger
F_{\Psi_\alpha}^{(n)} \right) \\ & & = \frac{1}{4} \sum_{k \in K^{(n)}
\atop \Psi,\alpha } {\rm Tr}\left( A_k A_k^\dagger F_{\Psi_\alpha}^{(n)}
\right) \; . \nonumber
\end{eqnarray}
The collision probability is split into contributions related to fixed
photon numbers arriving at Bob's detector in the same manner as the
disturbance measure to give
\begin{equation}
p_c^x  = \sum_{n=1}^\infty
\frac{p_{\rm rec}^{(n)}}{p_{\rm rec}} p_c^{(n)}
\end{equation}
with
\[
p_c^{(n)} := \sum_{k \in K^{(n)}, \Psi, \alpha}
\frac{1}{p^{(n)}_{\rm rec}}\frac{ p^2(\Psi_\alpha, k_\alpha)}{p(k_\alpha)}
\; .
\]
The basic idea is now to estimate the one-photon contributions to
these quantities and then to choose $w_D$ in such a way that the
optimal eavesdropping strategy will necessarily employ only one-photon
signals. To achieve this we will use the fact that multi-photon signals
lead unavoidably to ambiguous signals, that is $p_D^{(n)} \neq 0$ for
$n>2$ when using the passive detection option.

\subsection{ The one-photon contribution for  discarded errors}
We use the description of the general eavesdropping strategy to
calculate the one-photon contributions. We find with the help of
the identity $F_{\Psi_\alpha}^{(1)} = \frac{1}{2} \rho_{\Psi_\alpha}$
%\widetext
%\end{multicols}
\begin{eqnarray}
\label{pc1}
p_c^{(1)} &=& \frac{1}{8}\sum_{k \in K^{(1)}} \frac{1}{p_{\rm
rec}^{(1)}} \left\{ \frac{ {\rm Tr}^2\left(A_k \rho_{0_+} A_k^\dagger
\rho_{0_+}\right) +{\rm Tr}^2\left(A_k \rho_{1_+} A_k^\dagger
\rho_{1_+}\right)}{{\rm Tr}\left(A_k \rho_{0_+} A_k^\dagger
\rho_{0_+}\right) +{\rm Tr}\left(A_k \rho_{1_+} A_k^\dagger
\rho_{1_+}\right)} \right\} \\ & & +\frac{1}{8} \sum_{k' \in K'^{(1)}}
\frac{1}{p_{\rm rec}^{(1)}} \left\{ \frac{ {\rm Tr}^2\left(B_{k'}
\rho_{0_\times} B_{k'}^\dagger \rho_{0_\times}\right) +{\rm
Tr}^2\left(B_{k'} \rho_{1_\times} B_{k'}^\dagger
\rho_{1_\times}\right)}{{\rm Tr}\left(B_{k'} \rho_{0_\times}
B_{k'}^\dagger \rho_{0_\times}\right) +{\rm Tr}\left(B_{k'}
\rho_{1_\times} B_{k'}^\dagger \rho_{1_\times}\right)} \right\} \;
\nonumber
\end{eqnarray}
%\narrowtext 
%\begin{multicols}{2}
and with the relation between $p_{\rm rec}^{(1)}$ and
 $\overline{\epsilon}^{(1)}$ from eqn. (\ref{eps1def}), and $p_{\rm
 sif}^{(1)}=p_{\rm err}^{(1)} + p_{\rm rec}^{(1)}$ we find
\begin{equation}
\label{eN1}
  p^{(1)}_{\rm rec} =
 \frac{p^{(1)}_{\rm sif} }{1+ \overline{\epsilon}^{(1)}}
\end{equation}
together with the quantities
\begin{eqnarray}
\label{N1}
p_{\rm rec}^{(1)}&=&\frac{1}{8}\sum_{k \in K^{(1)} \atop \Psi, \alpha}
\left\{{\rm Tr}\left(A_k \rho_{\Psi_{\alpha}} A_k^\dagger
\rho_{\Psi_{\alpha}}\right) \right\} \\ \label{p1} p_{\rm sif}^{(1)}
&=& \frac{1}{4}\sum_{k \in
K^{(1)}}{\rm Tr}\left(A_k A_k^\dagger\right) \; .
\end{eqnarray}
The equations (\ref{pc1}--\ref{p1}) form the basis for the following
calculations. To start with, we decrease the number of free parameters
to a handful of real parameters, so that we can optimize Eve's
strategy to give an upper bound on $p_c^{(1)}$ as a function of
$\overline{\epsilon}^{(1)}$. To do so, we take a new look at the
complete positive mapping (\ref{cpmapping}). We define four vectors
${\bf A}_{00},{\bf A}_{10},{\bf A}_{01}, {\bf A}_{11}$ with the
components $k \in K^{(1)}$ given by
\begin{equation}
A_{\Psi, \Psi'}^k = \langle \Psi_+ | A_k | \Psi'_+ \rangle \; .
\end{equation}
These vectors are formed by the transition amplitudes from the signal
states to the one-photon detection states for each different
measurement outcome. They effectively describe not only the complete
channel between Alice and Bob but also the complete eavesdropping
strategy.  With these vectors we can simplify the notation of the
expectation values introducing vector products
$$
 \sum_{k \in K^{(1)}} {\rm Tr}\left(A_k \rho_{\Psi_+} A_k^\dagger
 \rho_{\Psi'_+} \right) = {\bf A}_{\Psi, \Psi'} {\bf A}_{\Psi, \Psi'}
 = \left| {\bf A}_{\Psi, \Psi'} \right|^2 \; .
$$
Similarly we can define vectors ${\bf B}_{00},{\bf B}_{10},{\bf
B}_{01}, {\bf B}_{11}$ and vectors ${\bf \tilde{B}}_{00},{\bf
\tilde{B}}_{10},{\bf \tilde{B}}_{01}, {\bf \tilde{B}}_{11}$ with
elements for $k' \in K'^{(1)}$
\begin{eqnarray}
B_{\Psi, \Psi'}^{k'} &=& \langle \Psi_+ | B_k' | \Psi'_+ \rangle \\
\tilde{B}_{\Psi, \Psi'}^{k'} &=& \langle \Psi_\times | B_k' |
\Psi'_\times \rangle\; .
\end{eqnarray}
These vectors are not independent. They are related by the identities 
\begin{eqnarray}
\label{vectorrelation}
{\bf \tilde{B}}_{00}&=& \frac{1}{2} \left({\bf B}_{00} - {\bf B}_{10}
- {\bf B}_{01} + {\bf B}_{11}\right) \\ {\bf \tilde{B}}_{01}&=&
\frac{1}{2} \left({\bf B}_{00} - {\bf B}_{10} + {\bf B}_{01} - {\bf
B}_{11}\right)\nonumber \\ {\bf \tilde{B}}_{10}&=& \frac{1}{2}
\left({\bf B}_{00} + {\bf B}_{10} - {\bf B}_{01} - {\bf
B}_{11}\right)\nonumber \\ {\bf \tilde{B}}_{11}&=& \frac{1}{2}
\left({\bf B}_{00} + {\bf B}_{10} + {\bf B}_{01} + {\bf
B}_{11}\right)\nonumber
\end{eqnarray}

The advantage of this description is that the value of any scalar
product of the vectors ${\bf B}_{\Psi,\Psi'}$ remains unchanged if the
${\bf B}_{\Psi,\Psi'}$'s are replaced by ${\bf A}_{\Psi,\Psi'}$'s
since (\ref{ABequality}) guarantees that
\begin{equation}
\label{dotproductrelation}
{\bf B}_{\Psi,\Psi'} {\bf B}_{\phi,\phi'} = {\bf A}_{\Psi,\Psi'} {\bf
A}_{\phi,\phi'} \; .
\end{equation}
The idea is now to estimate and reformulate the equations
(\ref{pc1}--\ref{p1}) in such a way that the new set of equations
involve only the four vectors ${\bf A}_{00},{\bf A}_{10},{\bf A}_{01},
{\bf A}_{11}$ and the quantities $\overline{\epsilon}^{(1)},
p_{\rm sif}^{(1)}$ and $p_{\rm rec}^{(1)}$. As a first step we find from
eqn. (\ref{pc1})
\begin{eqnarray}
p_c^{(1)} &=& \frac{1}{8 p_{\rm rec}^{(1)}} \sum_{k \in K^{(1)}}
\frac{(A^k_{00})^4 +(A^k_{11})^4 }{(A^k_{00})^2 +(A^k_{11})^2 }\\
 & &  +
\frac{1}{8 p_{\rm rec}^{(1)}} \sum_{k' \in K'^{(1)}}
\frac{(\tilde{B}^{k'}_{00})^4 +(\tilde{B}^{k'}_{11})^4
}{(\tilde{B}^{k'}_{00})^2 +(\tilde{B}^{k'}_{11})^2 } \; , \nonumber 
\end{eqnarray}
while equation (\ref{eN1}) remains unchanged
\begin{equation}
p^{(1)}_{\rm rec} = \frac{p^{(1)}_{\rm sif} }{1+
\overline{\epsilon}^{(1)}} \; .
\end{equation}
The definitions of $p_{\rm rec}^{(1)}$ and $\overline{\epsilon}^{(1)}$
simplify to
\begin{eqnarray}
\label{N1vector}
p_{\rm rec}^{(1)} & = & \frac{1}{8}\left(|{\bf A}_{00}|^2 +|{\bf
A}_{11}|^2 + |{\bf \tilde{B}}_{00}|^2 +|{\bf \tilde{B}}_{11}|^2\right)
\\
\label{p1vector}
p_{\rm sif}^{(1)} & = & \frac{1}{4} \left(|{\bf A}_{00}|^2 +|{\bf
A}_{11}|^2 + |{\bf A}_{01}|^2 +|{\bf A}_{10}|^2 \right)\; .
\end{eqnarray}
Next we use the Cauchy inequality as shown in appendix
\ref{cauchy} to estimate $p_c^{(1)}$ by an expression involving only
scalar products of the basic vectors.  With use of the definition of
$p_{\rm rec}^{(1)}$ this results in the expression
\begin{eqnarray}
\label{cauchyapplied}
\lefteqn{p_c^{(1)} \leq 1} \\
& & - \frac{1}{4 p_{\rm rec} } \frac{\left({\bf A}_{00}
{\bf A}_{11}\right)^2}{|{\bf A}_{00}|^2 +|{\bf A}_{11}|^2} -\frac{1}{4
p_{\rm rec} } \frac{\left({\bf \tilde{B}}_{00} {\bf
\tilde{B}}_{11}\right)^2}{|{\bf \tilde{B}}_{00}|^2 +|{\bf
\tilde{B}}_{11}|^2} \; . \nonumber
\end{eqnarray}
We find that there are actually only a few real quantities left. These
are $ |{\bf A}_{00}|, |{\bf A}_{11}|$, the angle $\phi_{00}^{11}$
between ${\bf A}_{00}$ and ${\bf A}_{11}$, $ |{\bf A}_{01}|^2+ |{\bf
A}_{10}|^2$, $ |{\bf A}_{01} + {\bf A}_{10}|^2$, $p_{\rm sif}$ and,
finally, $\overline{\epsilon}^{(1)}$. The normalization factor $p_{\rm
rec}^{(1)}$ can be immediately eliminated. As shown in appendix
\ref{optimisation} we can optimize $p_c^{(1)}$ and find the result
\begin{equation}
\label{pcdiscarded}
p_c^{(1)} \leq \left\{
\begin{array}{ll}
 \frac{1}{2} \left( 1 + 4 \overline{\epsilon}^{(1)} - 4 \left(
\overline{\epsilon}^{(1)} \right)^2 \right) & \mbox{for $
\overline{\epsilon}^{(1)} \leq 1/2$}
\\ 1 & \mbox{for $ \overline{\epsilon}^{(1)}  \geq 1/2$}
\end{array} \; .
\right.
\end{equation}
To compare this result with other results we introduce the error rate
$e$ in the sifted key as $e =
\frac{p_{\rm err}^{(1)}}{p_{\rm sif}^{(1)}}$ (so that
$\overline{\epsilon}^{(1)} = \frac{e}{1-e}$) and  we find
\begin{equation}
\label{pcdiscardedtrad}
 p_c^{(1)} \leq \frac{ 1 + 2 e - 7
e^2}{2 \left( 1 - e \right)^2} \; .
\end{equation}
This upper bound was given before in \cite{nl96b,nl97b} for the case
that Eve performed non-delayed measurements. Recently Slutsky et
al.~\cite{slutsky97a,slutsky98a} have found that this bound holds even
for the delayed case. My formulation of that proof shows that this
bound is valid not only for the one-photon contribution but can be
extended to include the full Hilbert space of optical fibers and
detectors accessible to Eve in real experiments.

From \cite{nl96b,nl97b,slutsky98a} we know that this bound is sharp
since the eavesdropping strategy achieving this bound is given
explicitly. It is a translucent attack. An important property of this
bound is that for a disturbance rate of $\overline{\epsilon}^{(1)}=
\frac{1}{2}$ (or error rate $e = \frac{1}{3}$) the eavesdropping
attempt is so successful that each bit of the sifted key originating
from this part of the eavesdropping strategy is known with unit
probability by Eve.

\subsection{ The one-photon contribution for  corrected errors}
If we correct errors without leaking knowledge about their position to
the eavesdropper, then the one photon contribution to the collision
probability is given by
\begin{eqnarray}
\label{pccorrected}
\lefteqn{p_c^{(1)} =}\\ & & \frac{1}{8 p_{\rm sif}^{(n)}} \sum_{k \in
K^{(1)}} \frac{ {\rm Tr}^2(\rho_{0_+} A_k^\dagger A_k) + {\rm
Tr}^2(\rho_{1_+} A_k^\dagger A_k)}{{\rm Tr}( A_k^\dagger A_k)}
\nonumber\\ & & +\frac{1}{8 p_{\rm sif}^{(n)}} \sum_{k' \in K'^{(1)}}
\frac{ {\rm Tr}^2(\rho_{0_\times} B_{k'}^\dagger B_{k'}) + {\rm
Tr}^2(\rho_{1_\times} B_{k'}^\dagger B_{k'})}{{\rm Tr}( B_{k'}^\dagger
B_{k'})} \; . \nonumber
\end{eqnarray}
(Note that $p_{\rm rec}^{(1)} =p_{\rm sif}^{(1)}$.) The disturbance
parameter coincides with the error rate $e^{(1)}$ in the sifted key
and is given by
\begin{equation}
\label{epscorr}
\overline{\epsilon}^{(1)} = \frac{p_{\rm err}^{(1)}}{p_{\rm sif}^{(1)}}
\end{equation}
with
\begin{eqnarray}
\label{psifcorr}
p_{\rm sif} & = & \frac{1}{4} \left(|{\bf A}_{00}|^2 +|{\bf A}_{11}|^2 +
|{\bf A}_{01}|^2 +|{\bf A}_{10}|^2 \right) \\ & = & \frac{1}{4}
\left(|{\bf \tilde{B}}_{00}|^2 +|{\bf \tilde{B}}_{11}|^2 + |{\bf
\tilde{B}}_{01}|^2 +|{\bf \tilde{B}}_{10}|^2 \right) \nonumber \\
\label{perrcorr}
p_{\rm err} & = &\\
& &  p_{\rm sif} - \frac{1}{8}\left(|{\bf A}_{00}|^2 +|{\bf
A}_{11}|^2 + |{\bf \tilde{B}}_{00}|^2 +|{\bf
\tilde{B}}_{11}|^2\right)\; . \nonumber
\end{eqnarray}
In appendix \ref{corrected} I show that the collision probability in
this case can be estimated by
\begin{equation}
\label{lepc}
p_c^{(1)} \leq \left\{
\begin{array}{ll}
\frac{1}{2} + 3 \overline{\epsilon}^{(1)} - 5
\left(\overline{\epsilon}^{(1)}\right)^2 & \mbox{for
$\overline{\epsilon}^{(1)} \leq 1/4$} \\ \frac{3}{4} +
\overline{\epsilon}^{(1)} - \left(\overline{\epsilon}^{(1)}\right)^2 &
\mbox{for $1/4 \leq \overline{\epsilon}^{(1)} \leq 1/2$} \\ 1 &
\mbox{for $1/2 \leq \overline{\epsilon}^{(1)}$}
\end{array}
\right. \; .
\end{equation}
This estimate is not necessarily sharp, but it is good enough for
practical purposes. It shows that $\tau_1 = 1$ for an error rate of
$\overline{\epsilon} = 1/2$, which corresponds to a strategy which
intercepts and stores all signals while random signals are resent. By
delaying the measurement of the signals Eve thus knows all signals
while causing a disturbance of $1/2$.

\subsection{One-photon contribution for corrected errors with leaked
error positions} If Alice and Bob use a bi-directional error
correction scheme then Eve will gain some knowledge about the
positions of the errors. She can therefore divide the signals into
subsets characterized by Eve's measurement outcome $k$, the
polarization basis $\alpha$ of the signal and the correctness of the
signal reception of Bob. We therefore need to introduce new operators
$C^k_{\Psi \Psi'}$ and $\tilde{D}^k_{\Psi \Psi'}$ to describe the
eavesdropping strategy applied to incorrectly received signals. They
are formed analogous to $A^k_{\Psi \Psi'}$ and $\tilde{B}^k_{\Psi
\Psi'}$ respectively.  Then the one-photon contribution towards the
collision probability is given by
\begin{eqnarray}
\label{pccorrpos}
p_c^{(1)}& =& \frac{1}{8 p_{\rm sif}^{(1)}} \sum_{k \in K^{(1)}}
\frac{(A^k_{00})^4 +(A^k_{11})^4 }{(A^k_{00})^2 +(A^k_{11})^2 }\\
 & &  +
\frac{1}{8 p_{\rm sif}^{(1)}} \sum_{k' \in K'^{(1)}}
\frac{(\tilde{B}^{k'}_{00})^4 +(\tilde{B}^{k'}_{11})^4
}{(\tilde{B}^{k'}_{00})^2 +(\tilde{B}^{k'}_{11})^2 } \\
 & & +
\frac{1}{8 p_{\rm sif}^{(1)}} \sum_{k \in K^{(1)}} \frac{(C^k_{01})^4
+(C^k_{10})^4 }{(C^k_{01})^2 +(C^k_{10})^2 } \\
 & &  + \frac{1}{8
p_{\rm sif}^{(1)}} \sum_{k' \in K'^{(1)}} \frac{(\tilde{D}^{k'}_{01})^4
+(\tilde{D}^{k'}_{10})^4 }{(\tilde{D}^{k'}_{01})^2
+(\tilde{D}^{k'}_{10})^2 } \; . \nonumber
\end{eqnarray}
The disturbance $\overline{\epsilon}^{(1)}$, $ p_{\rm sif}$ and
$p_{\rm err}$ are defined as in eqns. (\ref{epscorr}) to
(\ref{perrcorr}) where we note that within scalar products like
equation (\ref{dotproductrelation}) the vectors ${\bf C}$ (${\bf
\tilde{D}}$) can be replaced by ${\bf A}$ (${\bf \tilde{B}}$). In
appendix \ref{correctedleaked} I show that
\begin{equation}
p_c^{(1)} \leq \left\{
\begin{array}{ll}
\frac{1}{2} + 2 \overline{\epsilon}^{(1)} - 2
\left(\overline{\epsilon}^{(1)}\right)^2 & \mbox{for
$\overline{\epsilon}^{(1)} \leq 1/2$} \\ 1 & \mbox{for $1/2 \leq
\overline{\epsilon}^{(1)}$}
\end{array}
\right. \; .
\end{equation}
As it is the case if the error positions are not known to Eve, this
estimate is not necessarily sharp. This is due to the use of the
Cauchy inequality during the estimation. It shows a behavior analogous
to that of equation (\ref{lepc}) that for an error rate of $e=1/2$
(and disturbance rate $\overline{\epsilon}= 1/2$) we find
$\tau_1(1/2)=1$ which
means that Eve knows the whole key.

\subsection{Multi-photon signals between Eve and Bob}
\label{multisection}
To deal with multi-photon signals we have to pick a detection
model. We will concentrate here on the passive detection scheme to
choose $w_D$ such that it is disadvantageous for Eve to use
multi-photon signals. In my thesis \cite{nl96b} I have shown that even
for active switching between two polarization analyzer with different
polarization orientation one can show security against eavesdropping
strategies employing multi-photon signals.

The crucial observation for the passive detection unit is that sending
multi-photon signals will invariably cause the outcome associated with
$F_D$ to appear with a finite probability. This means that we can
choose the weight factor $w_D$ such that $ \overline{\epsilon}^{(n)} >
\overline{\epsilon}^{(1)}$ holds for $n \geq 2$.  As a consequence the
optimal eavesdropping strategy will employ only single-photon
signals. The contribution of ambiguous signals to the disturbance
parameter $\overline{\epsilon}^{(n)} $ for discarded errors is bounded
by a rough estimate obtained with help of eqn.~(\ref{passivpom}) by
omission of suitable positive terms in the expression for $F_D$
\begin{eqnarray}
\frac{p_D^{(n)}}{p_{\rm rec}^{(n)}} & = & \frac{ \frac{1}{4} \sum_{k
\in K^{(n)}\atop \Psi,\alpha} {\rm Tr}( A_k \rho_{\Psi_\alpha} A_k^\dagger
F_D)}{\frac{1}{4} \sum_{k \in K^{(n)} \atop \Psi,\alpha } {\rm Tr}\left(
A_k \rho_{\Psi_\alpha} A_k^\dagger F_{\Psi_\alpha}^{(n)} \right)} \\ &
\geq & \frac{\frac{1}{4} \sum_{k \in K^{(n)} \atop \Psi,\alpha
}\left(\frac{1}{2}-2^{-n} \right) {\rm Tr}\left(
A_k\rho_{\Psi_\alpha} A_k^\dagger E_{\Psi_\alpha}^{(n)} \right)}{
2^{-n}\frac{1}{4} \sum_{k \in K^{(n)} \atop \Psi,\alpha } {\rm Tr}\left(
A_k \rho_{\Psi_\alpha} A_k^\dagger E_{\Psi_\alpha}^{(n)} \right)}
\nonumber \\ & = & \frac{ \left(\frac{1}{2}-2^{-n} \right) }{ 2^{-n} }
\geq 1 \; .\nonumber
\end{eqnarray}
The contribution of ambiguous signals to the disturbance parameter
$\overline{\epsilon}^{(n)} $ for corrected errors is bounded in the
same way as
\begin{eqnarray}
\frac{p_D^{(n)}}{p_{\rm sif}^{(n)}} & = & \frac{ \frac{1}{4} \sum_{k
\in K^{(n)}\atop \Psi,\alpha} {\rm Tr}( A_k \rho_{\Psi_\alpha}
A_k^\dagger F_D)}{\frac{1}{4} \sum_{k \in K^{(n)} \atop
\Psi,\Psi',\alpha } {\rm Tr}\left( A_k \rho_{\Psi_\alpha} A_k^\dagger
F_{\Psi'_\alpha}^{(n)} \right)} \\ & \geq & \frac{\frac{1}{4} \sum_{k
\in K^{(n)} \atop \Psi,\Psi',\alpha }\left(\frac{1}{2}-2^{-n} \right)
{\rm Tr}\left( A_k\rho_{\Psi_\alpha} A_k^\dagger
E_{\Psi'_\alpha}^{(n)} \right)} { 2^{-n}\frac{1}{4} \sum_{k \in
K^{(n)} \atop \Psi,\Psi',\alpha } {\rm Tr}\left( A_k
\rho_{\Psi_\alpha} A_k^\dagger E_{\Psi'_\alpha}^{(n)}
\right)}\nonumber \\ & = & \frac{\frac{1}{4} \left(\frac{1}{2} -2^{-n}
\right) }{ 2^{-n}\frac{1}{4} } \geq 1 \; . \nonumber
\end{eqnarray}
One can find lower values of $w_D$ estimating the expression for
$\overline{\epsilon}^{(n)}$ as a whole including the errors in the
sifted key. However, the values found here serve our purposes well
enough.

\begin{figure}
\centerline{\psfig{width=8cm,file=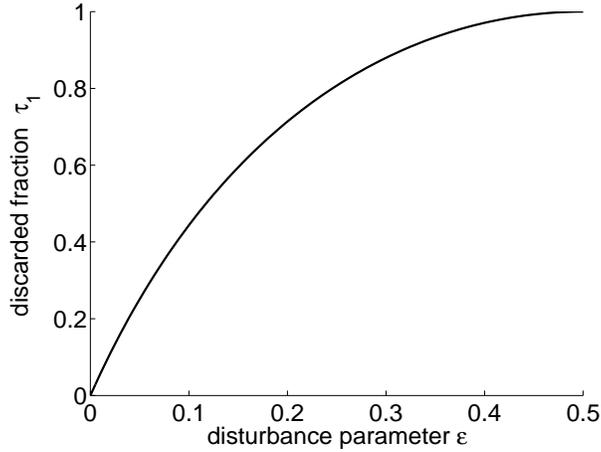}}
\caption{\label{displot} The fraction $\tau_1$ has to be discarded
during privacy amplification as a function of the disturbance per
correctly received element of the generalized sifted key if errors are
discarded. This result is a sharp estimate in the sense that Eve can
reach the level of collision probability on which the estimate is
based. }
\end{figure}

\begin{figure}
\centerline{\psfig{width=8cm,file=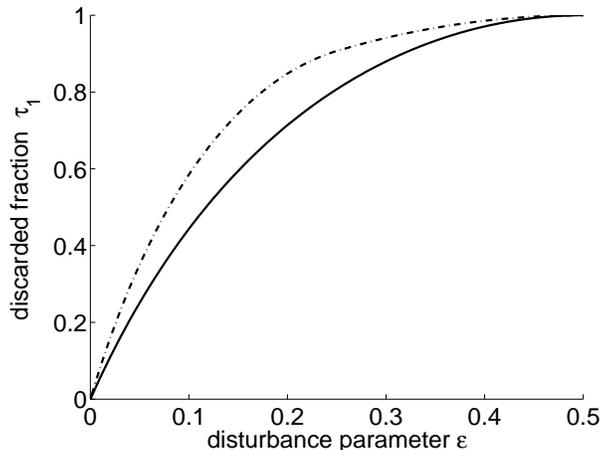}}
\caption{\label{corrplot} The fraction $\tau_1$ has to be discarded
during privacy amplification as a function of the disturbance per
element of the generalized sifted key if one corrects  errors. If no
information about the position of errors leaked to the eavesdropper,
we find for $\tau_1$ the dash-dotted curve, for leaked error positions
we find the solid curve.}
\end{figure}

For correcting and for discarding errors, we find that a disturbance
parameter $\overline{\epsilon}= 1/2$ means that Eve knows the whole
key using one-photon signals.  Therefore, if we choose $w_D =
\frac{1}{2}$ we obtain $\overline{\epsilon}^{(n)} \geq w_D
\frac{p_D^{(n)}}{p_{\rm rec}^{(n)}} \geq \frac{1}{2}$ and
$\overline{\epsilon}^{(n)} \geq w_D \frac{p_D^{(n)}}{p_{\rm
sif}^{(n)}} \geq \frac{1}{2}$ respectively and can bound the collision
probability, taking into account the possibility of multi-photon
signals, for discarded errors by
\begin{equation}
\label{taudis}
\tau_1(\overline{\epsilon}) \leq \left\{
\begin{array}{ll}
 \log \left( 1 + 4 \overline{\epsilon} - 4 
\overline{\epsilon}^2 \right) & \mbox{for
$\overline{\epsilon} \leq 1/2$} \\ 1 & \mbox{for $ 1/2 \leq
\overline{\epsilon} $}
\end{array}
\right. \;,
\end{equation}
for corrected errors without leaked error position by
\begin{equation}
\label{taucorrnoleak}
\tau_1(\overline{\epsilon}) \leq \left\{
\begin{array}{ll}
\log \left(1 + 6 \overline{\epsilon} - 10 \overline{\epsilon}^2\right)
& \mbox{for $\overline{\epsilon} \leq 1/4$} \\ \log \left(\frac{3}{2}
+2 \overline{\epsilon} - 2 \overline{\epsilon}^2\right) & \mbox{for
$1/4 \leq\overline{\epsilon} \leq 1/2$} \\ 1 & \mbox{for $1/2
\leq\overline{\epsilon}$}
\end{array}
\right. \; ,
\end{equation}
and for corrected errors with leaked error positions by
\begin{equation}
\label{taucorrleak}
\tau_1(\overline{\epsilon}) \leq \left\{
\begin{array}{ll}
 \log \left( 1 + 4 \overline{\epsilon} - 4 
\overline{\epsilon}^2 \right) & \mbox{for
$\overline{\epsilon} \leq 1/2$} \\ 1 & \mbox{for $ 1/2 \leq
\overline{\epsilon} $}
\end{array}
\right. \; .
\end{equation}
The results for $\tau_1$ are shown in figure \ref{displot} and
\ref{corrplot} respectively. It should be noted again, that the value
of the disturbance parameter changes depending on the intention to
correct the errors.  For other detector models these results hold as
well as long as we can show that for them the condition
$\overline{\epsilon}^{(n)} \geq 1/2$ for $n \geq 2$ holds. This
condition can be readily satisfied if $p_D^{(n)} / p_{\rm rec}^{(n)}
\geq \mu$ for some $\mu > 0$ and $n \geq 2$ by choosing $w_d = 1/(2
\mu)$.  For experiments with negligible numbers of ambiguous results
we can approximate the disturbance $\overline{\epsilon}$ by a function
of $e = \frac{p_{\rm err}}{p_{\rm sif}}$ as the traditional error rate
in the sifted key. In the case of discarding errors this approximation
is $\overline{\epsilon} \approx \frac{e}{1-e}$ while for corrected
keys it is $\overline{\epsilon} \approx e$.

Since we can not give an estimate for $\overline{\epsilon}$ from
measured quantities the case of discarded errors, we concentrate on
reconciliation methods which correct errors. From the results of this
section we see that this is the better methods anyway, since
discarding errors leads to a smaller $n_{\rm rec}$ than correcting
errors. This number would have to be reduced further during privacy
amplification than in the case of corrected errors, as can be seen by
comparison of the estimates for $\tau_1$ as a function of
$e$. Therefore the final key will be shorter and with that the
protocol less efficient.

From the estimates we find that the direct estimate for $\tau_1$ gives
higher values if the information about error positions has not leaked
to the eavesdropper during reconciliation.  We can regard the
information of error positions as {\it spoiling information}
\cite{bennett95a} and thus use the estimate (\ref{taucorrleak}) even
in the case of uni-lateral error correction. Spoiling information is
any information which increases Eve's Shannon information but
decreases her expected collision probability on the key leading to a
decreased value of $\tau_1$.  We conclude that from the point of
privacy amplification and reconciliation, the best known way to give a
high rate of secure bits would be to use bi-lateral reconciliation
methods.

\section{Analysis of the efficiency of key growing}
\label{analysis}
The process of quantum key growing depends on physical parameters and
on the security parameters of the final key. In this section we will
bring together the essential formulas about the security statements
concerning an accepted key and about the average key growing rate we
can expect. This analysis is presented only for error correction
reconciliation methods.

\subsection{Security needs}
The first thing a potential user has to fix is the tolerated change of
Shannon entropy $\Delta_{\rm tol}$ an eavesdropper might obtain on the
key without posing a security hazard to the application in mind. Since
this limit can not be guaranteed with absolute certainty, the user has
to limit the tolerated probability $\alpha_{\rm tol}$ that Eve's
knowledge exceeds $\Delta_{\rm tol}$. Authentication may fail to
detect errors leaving Alice and Bob with a key neither safe nor
shared. The tolerated probability for this has to be specified as
$\gamma_{\rm tol}$.

Given $I_{\rm tol}$, $\alpha_{\rm tol}$ and $\gamma_{\rm tol}$ and
having in view a particular physical implementation of the quantum
channel, Alice and Bob fix a value of the tolerated disturbance
$\overline{\epsilon}_{\rm max}$ and of the security bits $n_S$ used in
privacy amplification, as well as the length $n_{\rm sif}$ of the
sifted key and the number of secure bits $N_{\rm aut}$ used for
authentication such that for an accepted key
the security target set by $I_{\rm tol},\alpha_{\rm tol} $ and
$\gamma_{\rm tol}$ is met and that the rate of secure bits generated,
given below, is optimized.

\subsection{Security statement}
The following security statement holds if the key growing is performed
by extracting a key of length
\begin{equation}
n_{\rm fin} = n_{\rm sif}\left[1- \tau_1(\overline{\epsilon}_{\rm
max})\right]-n_S
\end{equation}
from the reconciled key during privacy amplification. Here $\tau_1$ is
given by by the functional dependence of equations
(\ref{taucorrnoleak}) and (\ref{taucorrleak}) respectively.  From the
previous calculations we find that the bits generated in a run of the
key growing process are secure in the sense that Eve achieves a change
of Shannon entropy on the accepted key of less than $\Delta_{\rm tol}$
with probability $\alpha$. The contributions to $\alpha$ are the
probability of failure of the estimation of the average disturbance
given by $\alpha_1$ in equation (\ref{alpha1}), the probability to
estimate the Shannon information in a specific run from the average
information, given by $\alpha_2$ in equation (\ref{alpha2}) and the
probability of faked authentication, given by $\alpha_3$ in equation
(\ref{alpha3}). Since all those quantities are expected to be small,
the estimate
\begin{eqnarray}
\alpha &\leq& \alpha_1 + \alpha_2 + \alpha_3\\ & & = \exp(-2 n_{\rm
sif} \delta^2) + \frac{\ln (2^{-n_S} + 1)}{\Delta_{\rm tol}} +
2^{-N_{\rm aut}+1} \nonumber\\ & \approx& \exp(-2 n_{\rm sif}
\delta^2)+ \frac{2^{-n_S}}{\Delta_{\rm tol} \ln 2} + 2^{-N_{\rm
aut}+1} \nonumber
\end{eqnarray}
with $\delta =\overline{\epsilon}_{\rm max}- \epsilon_{\rm meas}$ is
sufficient for practical purposes.

The failure to establish a key in a specific run is due to the failure
of authentication. Here two contributions can be distinguished. One is
the failure of reconciliation, which happens with probability
$\beta_1$, the other is the failure to reach the target of
$\alpha_{\rm tol}$ in that run, which is signaled by making the
authentication fail. This happens with a probability $\beta_2$. In the
design of the set-up and the choice of parameters we would need to
estimate $\beta$ so that at least in the absence of an eavesdropper we
will find a net gain of secure bits according to the formula given
below. Miscalculation of $\beta$ does not affect the security of the
key, it only affects the efficiency of key generation. We omit
therefore detailed examinations of values for $\beta$.

The last quantity concerning the security of the key is $\gamma$,
which is the probability that authentication succeeds although Alice
and Bob do not share a key. This probability can be estimated by
$\gamma = 2^{-N_{\rm aut}+1}$.

\subsection{Gain}
\begin{figure}
\centerline{\psfig{width=8cm,file=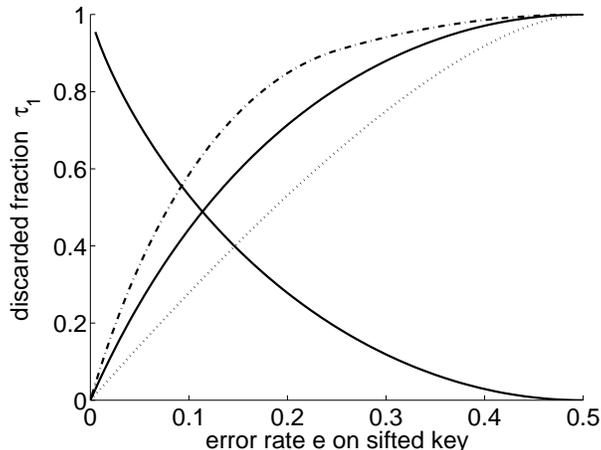}}
\caption{\label{limitplot} Shortening during privacy amplification,
represented by $\tau_1$ (uni-lateral scenario in dash-dotted curve,
bi-lateral scenario as solid curve), in balance with the loss during
reconciliation, represented by $I_{AB}$ (falling solid line). The
intersections between two lines limits the tolerable error rate in the
generalized sifted key in the case of corrected errors. A lower limit
of potentially improved bounds for $\tau_1$ is $I_{AE}$ (dotted
line).}
\end{figure}

In the previous subsection we described the influence of the chosen
basic parameters on the acceptance and security of a run of key
growing.  Since we need secret bits as an input for the key generation
we have to make sure that on average we will gain more secret bits
than we put in. The important quantities are here the success
probability $p_{\rm succ} = 1-\beta$ that a run of the key expansion
leads to accepted new secure bits, the number $N_{\rm out} = n_{\rm
rec}\; [1- \tau_1(\overline{\epsilon}_{\rm max})]-n_S $ of secret bits
gained in that instance and the average number $\overline{N}_{\rm
in}=\overline{N}_{\rm rec}+N_{\rm aut}$ of input secret bits.  Then
the condition for an overall gain on average is to have a positive
value of $\overline{N}_{\rm gain} = p_{\rm succ} N_{\rm out} -
\overline{N}_{\rm in}$ resulting in
\begin{eqnarray}
\label{gaineqn}
\overline{N}_{\rm gain} &=& (1- \beta) \left\{n_{\rm sif}\left[1-
 \tau_1(\overline{\epsilon}_{\rm max})\right]-n_S \right\} \\
& & -N_{\rm aut}-N_{\rm rec}\; . \nonumber 
\end{eqnarray}
 To explore the implications of this condition we go to the limit of
large sample sizes. Then we can neglect the number of secret bits used
for authentication and and the safety parameter $n_S$. The remaining
contribution of $\overline{N}_{\rm in}$ now comes from the error
correction part. For ideal error correction we can set $\beta = 0$ and
can use the Shannon limit which gives $\overline{N}_{\rm in} = n_{\rm
sif} (1- I_{AB}(\epsilon_{\rm meas}))$ with the Shannon information
shared between Alice and Bob given by
\begin{eqnarray}
\lefteqn{I_{AB}(\epsilon_{\rm meas}) =}\\
& &  1 + \epsilon_{\rm meas} \log \epsilon_{\rm meas} +
(1-\epsilon_{\rm meas}) \log (1-\epsilon_{\rm meas}) \; . \nonumber
\end{eqnarray}
With these preparations we find $$N_{\rm gain} = n_{\rm sif} \left[1-
\tau_1(\epsilon_{\rm meas})\right]- n_{\rm sif} (1-
I_{AB}(\epsilon_{\rm meas}))\; .$$ In the limit of $n_{\rm sif} \to
\infty$ we can assume that $\delta \to 0$ still satisfies any
confidence limits put on $\alpha$. Therefore the condition
$\overline{N}_{\rm gain} \geq 0$ is now equivalent to
\begin{equation}
I_{AB}(\epsilon_{\rm meas}) \geq \tau_1(\epsilon_{\rm meas}) \; .
\end{equation}
As we see from figure \ref{limitplot} this means that the protocol in
the presented form will be able to grow secret keys only for set-ups
operating at an error rate of less than $11.5\%$ for error
correction. However, making use of the concept of spoiling information
and of improved estimates of $p_c^{(1)}$ might result in lower
estimates for $\tau_1$. A lower bound is, however, the Shannon
information $I_{AE}$ shared by Alice and Eve in this scenario. Fuchs
et al.~give in \cite{fuchs97a} a sharp bound for $I_{AE}$, which is
shown in figure \ref{limitplot} as dotted line.  The difference
between $\tau_1$ and $I_{AE}$ represent the average gain $G$ in a run
of the key growing protocol in the limit of ideal error correction and
infinite sample sizes. The gain
\begin{equation}
G = I_{AB}(\epsilon_{\rm meas}) - \tau_1(\epsilon_{\rm meas})
\end{equation}
gives the length of the final key as a fraction of the generalized
sifted key.

\section{Concluding Remarks}
In this paper I have given estimates needed in quantum cryptography
which are closely oriented towards practical experiments. I do not
deal with security against all possible attacks in quantum mechanics,
but I deal with all attacks on individual signals. This allows me to
include issues related to practical implementation of quantum
cryptography which still can not be treated in the general
scenario. One of these issues is the question of signals which, for
example, triggered simultaneously two detectors monitoring orthogonal
polarization modes. (This is the question of multi-photon signals
resent by Eve, leading to ambiguous signals.) The other important
question is that of an efficient key reconciliation prior to privacy
amplification. As seen in this paper it is possible to use the
efficient bi-lateral error correction scheme of Brassard and Salvail
\cite{brassard93a} without compromising security.

In the statistical analysis I showed that it is possible to limit in
this scenario the knowledge of the eavesdropper on the final key in a
individual realization from {\it measured quantities} for parameters
which seem to be reachable in experiments. As measure of the
eavesdropper's knowledge I used the change between a-priori and
a-posteriori Shannon entropy associated with the corresponding
probability distributions over all possible keys from Eve's point of
view. One has to take into account that single photon signals states
are not used in today's experiments. However, this theory can be
extended to signal states containing multi-photon components. A first
approach for that is to estimate $p_c^x=1$ for each bit of the
reconciled key on which Eve could have performed successfully a
splitting operation with subsequent delayed measurement. Denote by
$n_m$ the total number of these bits, then we  need to reduce the
key during privacy amplification by
\begin{equation}
\tau_1^{(mult)}(\overline{\epsilon}) = \frac{n_m}{n_{\rm rec}} +
\left(1-\frac{n_m}{n_{\rm rec}}\right) \tau_1\left(\overline{\epsilon}
\frac{n_{\rm rec}}{n_{\rm rec}-n_m}\right) \; .
\end{equation} 
The statistics, however, becomes more complicated this way and it
seems to be better to include the dim coherent states directly as
signal states and to solve the problem in a clean way. Work in that
direction is currently under progress.

The estimates for $\tau_1$ are not necessarily sharp in the case of
error correction, and even in the case of discarding errors this limit
could be lowered using spoiling information \cite{bennett95a}.
However, the possible improvement of efficiency of the key growing
process is limited and this fine-tuning might be postponed until
the experimental relevant situation for dim coherent signal states is
solved.

\section*{Acknowledgments}
I would like to thank Miloslav Du\v{s}ek, Richard Hughes, Paul
Townsend and the participants of the 1997 workshop on quantum
information at the Institute for Scientific Interchange (Italy) for
discussions and Steven van Enk for helpful critical comments on the
manuscript. For fincancial support I would like to thank Elsag--Bailey
and the Academy of Finland.  The foundations to this article were laid
while I did research for my PhD thesis under supervision and support
of Steve Barnett.

\appendix
\section{Cauchy inequality}
\label{cauchy}
In this appendix we prove the inequality (\ref{cauchyapplied})
starting from the expression
\begin{eqnarray}
\label{intermediate2}
p_c^{(1)} &=& \frac{1}{8 p_{\rm rec}^{(1)}} \sum_{k \in K^{(1)}}
\frac{(A^k_{00})^4 +(A^k_{11})^4 }{(A^k_{00})^2 +(A^k_{11})^2 }\\
& &  +
\frac{1}{8 p_{\rm rec}^{(1)}} \sum_{k' \in K'^{(1)}}
\frac{(\tilde{B}^{k'}_{00})^4 +(\tilde{B}^{k'}_{11})^4
}{(\tilde{B}^{k'}_{00})^2 +(\tilde{B}^{k'}_{11})^2 } \; . \nonumber
\end{eqnarray}
We rewrite the first sum as 
\begin{equation}
\label{intermediate1}
\sum_k \left((A^k_{00})^2 +(A^k_{11})^2- 2 \frac{ \left(A^k_{00}
A^k_{11} \right)^2}{(A^k_{00})^2 +(A^k_{11})^2 }\right) \; 
\end{equation}
and use the Cauchy inequality, given as
\begin{equation}
\left(\sum_k x_k y_k \right)^2 \leq \left( \sum_k x_k^2\right) \left(
\sum_k y_k^2 \right)
\end{equation}
or
\begin{equation}
 \sum_k x_k^2 \geq \frac{\left(\sum_k x_k y_k \right)^2}{ \sum_k y_k^2
 \; .  }
\end{equation}
 We set $x_k = \frac{ \left(A^k_{00} A^k_{11}
\right)}{\sqrt{(A^k_{00})^2 +(A^k_{11})^2 }}$ and $ y_k =
\sqrt{(A^k_{00})^2 +(A^k_{11})^2 }$ to obtain the inequality 
\begin{eqnarray}
\lefteqn{\sum_k\frac{(A^k_{00})^4 +(A^k_{11})^4 }{(A^k_{00})^2
+(A^k_{11})^2 } \leq}\\ & & \sum_k \left((A^k_{00})^2 +(A^k_{11})^2
\right) - 2 \frac{ \left(\sum_k A^k_{00} A^k_{11}
\right)^2}{\sum_k(A^k_{00})^2 +(A^k_{11})^2 } \; .  \nonumber
\end{eqnarray}
This can be used to estimate the first part in (\ref{intermediate2})
while the second part can be estimated similarly so that, with the
help of eqn.~(\ref{N1vector}), we find the result
\begin{eqnarray}
\lefteqn{p_c^{(1)} \leq 1}\\
& & - \frac{1}{4 p_{\rm rec} } \frac{\left({\bf A}_{00} {\bf
A}_{11}\right)^2}{|{\bf A}_{00}|^2 +|{\bf A}_{11}|^2} -\frac{1}{4
p_{\rm rec} } \frac{\left({\bf \tilde{B}}_{00} {\bf
\tilde{B}}_{11}\right)^2}{|{\bf \tilde{B}}_{00}|^2 +|{\bf
\tilde{B}}_{11}|^2} \; . \nonumber
\end{eqnarray}

\section{Maximizing $p_c^{(1)}$ for discarded errors}
\label{optimisation}
To optimize the expression (\ref{cauchyapplied}) we first note that we
can assume that $|{\bf A}_{00}| = |{\bf A}_{11}|$. If Eve starts with
a strategy defined by operators $A_k$ not satisfying this condition,
then she could use the A-operators $\overline{A}_k = \left(
\begin{array}{cc}
0& 1\\1& 0
 \end{array}\right) A_k\left(
\begin{array}{cc}
0& 1\\1& 0 \end{array}\right)$ without a change in the obtained
 collision probability or disturbance. When we combine the two
 strategies we find that the resulting vectors satisfy $|{\bf A}_{00}|
 = |{\bf A}_{11}|$ and $|{\bf A}_{01}| = |{\bf A}_{10}|$. This then
 gives the estimate $|{\bf A}_{01} + {\bf A}_{10}|^2\leq 4 |{\bf
 A}_{01}|^2$.  Another observation is that we can always choose $|{\bf
 A}_{00}| + |{\bf A}_{11}| \geq |{\bf \tilde{B}}_{00}| + |{\bf
 \tilde{B}}_{11}|$ which means that there are less or equal errors in
 the sifted key coming from the use of the polarization basis '+' than
 from the basis '$\times$'. This can be always satisfied, since both
 polarization basis could be interchanged. Using $|{\bf A}_{00}| =
 |{\bf A}_{11}|$ and the definition of $|{\bf \tilde{B}}_{00}|$ and
 $|{\bf \tilde{B}}_{11}|$ this results in $2 |{\bf A}_{00}|^2 (1- \cos
 \phi_{00}^{11} ) \geq |{\bf A}_{01} + {\bf A}_{10}|^2$ with the angle
 $\phi_{00}^{11}$ between ${\bf A}_{00}$ and ${\bf A}_{11}$.

The three relevant relations now become after elimination of $p_{\rm
rec}^{(1)}$ according to (\ref{eN1}) and the use of the relations
(\ref{vectorrelation}) 
%\widetext
%\end{multicols}
\begin{eqnarray}
p_c^{(1)}& \leq& 1 - \frac{(1 + \overline{\epsilon}^{(1)})|{\bf
A}_{00}|^2 (\cos \phi_{00}^{11})^2}{8 p_{\rm sif} } - \frac{(1 +
\overline{\epsilon}^{(1)}) \left(2 |{\bf A}_{00}|^2 (1+\cos
\phi_{00}^{11})- |{\bf A}_{01}+{\bf A}_{10}|^2 \right)^2}{32 p_{\rm sif}
\left(2 |{\bf A}_{00}|^2 (1+\cos \phi_{00}^{11})+ |{\bf A}_{01}+{\bf
A}_{10}|^2 \right)} \\ \label{middleeq} \frac{ p_{\rm sif}}{(1 +
\overline{\epsilon}^{(1)})} & = &\frac{1}{8} \left( |{\bf A}_{00}|^2
(3+\cos \phi_{00}^{11})+ \frac{1}{2}|{\bf A}_{01}+{\bf
A}_{10}|^2\right) \\ p_{\rm sif} & = & \frac{1}{2} \left(|{\bf A}_{00}|^2
+ |{\bf A}_{01}|^2\right)
\end{eqnarray}
Our next step is to show that we can estimate the optimal value of
$p_c^{(1)}$ by replacing $|{\bf A}_{01}+{\bf A}_{10}|^2$ by $4 |{\bf
A}_{01}|^2$.  To see that we observe that this would allow to decrease
$(1 + \overline{\epsilon}^{(1)})$ by eqn (\ref{middleeq}), meaning a
lower error rate. At the same time $p_c^{(1)}$ grows indirectly from
the falling value of $(1 + \overline{\epsilon}^{(1)})$ and directly,
since $\frac{d}{dD}p_c^{(1)} \geq 0$ with $D := |{\bf A}_{01}+{\bf
A}_{10}|^2$. To prove the last point we calculate
\begin{eqnarray}
\frac{d}{dD}p_c^{(1)} &=& \frac{(1 + \overline{\epsilon}^{(1)}){\cal
A}}{32 p_{\rm sif} \left( 2 |{\bf A}_{00}|^2 + D + 2 |{\bf
A}_{00}|^2\cos \phi_{00}^{11}\right)^2} \\ {\cal A} & = & 12 |{\bf
A}_{00}|^4 - 4 |{\bf A}_{00}|^2 D - D^2+ 24|{\bf A}_{00}|^4 \cos
\phi_{00}^{11} - 4 |{\bf A}_{00}|^2 D\cos \phi_{00}^{11} + 12 |{\bf
A}_{00}|^4 (\cos \phi_{00}^{11})^2 \; .
\end{eqnarray}
This is positive, if ${ \cal A} $ is positive. This is, indeed, the
case since
\begin{equation}
\frac{d}{dD} {\cal A}=- 4 |{\bf A}_{00}|^2- 2 D- 4 |{\bf A}_{00}|^2
\cos \phi_{00}^{11} \leq 0
\end{equation}
allows us to evaluate ${ \cal A} $ at the maximal value of $D_{\rm
max} = 2 |{\bf A}_{00}|^2 (1- \cos \phi_{00}^{11} ) $ where it gives
zero. This proves that ${\cal A} \geq 0 $ and with that
$\frac{d}{dD}p_c^{(1)}\geq 0 $. Therefore, three relevant equations
become
\begin{eqnarray}
\label{pc3} p_c^{(1)}& \leq& 1 -
 \frac{(1 + \overline{\epsilon}^{(1)})|{\bf A}_{00}|^2 (\cos
 \phi_{00}^{11})^2}{8 p_{\rm sif} } - \frac{(1 +
 \overline{\epsilon}^{(1)}) \left( |{\bf A}_{00}|^2 (1+\cos
 \phi_{00}^{11})-2 |{\bf A}_{01}|^2 \right)^2}{16 p_{\rm sif} \left(
 |{\bf A}_{00}|^2 (1+\cos \phi_{00}^{11})+ 2 |{\bf A}_{01}|^2 \right)}
 \\
\label{e3} \frac{ p_{\rm sif}}{(1 + \overline{\epsilon}^{(1)})}  & = &
\frac{1}{8} \left( |{\bf A}_{00}|^2 (3+\cos \phi_{00}^{11})+2 |{\bf
A}_{01}|^2 \right) \\
\label{p3}
 p_{\rm sif} & = &\frac{1}{2} \left( |{\bf A}_{00}|^2 +|{\bf
 A}_{01}|^2\right)
\end{eqnarray}
%\narrowtext
We solve (\ref{e3}) and (\ref{p3}) for $|{\bf A}_{01}|$ and $\cos
\phi_{00}^{11}$ and insert these into (\ref{pc3}). The maximum over
$|{\bf A}_{00}|$ is then taken  and we find
\begin{equation}
p_c^{(1)} \leq \frac{1}{2} \left( 1 + 4 \overline{\epsilon}^{(1)} - 4
\left( \overline{\epsilon}^{(1)} \right)^2 \right) \; .
\end{equation}
The strategy resulting in this collision probability is described by
\begin{eqnarray}
|{\bf A}_{00}|^2 & = & |{\bf A}_{11}|^2 = \frac{2 p_{\rm sif}
}{1+\overline{\epsilon}^{(1)}} \\ |{\bf A}_{01}|^2 & = & |{\bf
A}_{10}|^2 = \frac{2 p_{\rm sif}
\overline{\epsilon}^{(1)}}{1+\overline{\epsilon}^{(1)} }\\ \cos
\phi_{00}^{11} & = &1-2\overline{\epsilon}^{(1)} \\ \cos
\phi_{01}^{10} & = & 1 \; .
\end{eqnarray}
In the derivation we have chosen $2 |{\bf A}_{00}|^2 (1- \cos
\phi_{00}^{11} ) \geq |{\bf A}_{01} + {\bf A}_{10}|^2$ and find the
optimal solution respects this choice for $ \overline{\epsilon}^{(1)}
\leq \frac{1}{2}$.  For $ \overline{\epsilon}^{(1)} = \frac{1}{2}$ we
find $p_c^{(1)} = 1$ so that we conclude that
\begin{equation}
p_c^{(1)} \leq \left\{
\begin{array}{ll}
 \frac{1}{2} \left( 1 + 4 \overline{\epsilon}^{(1)} - 4 \left(
\overline{\epsilon}^{(1)} \right)^2 \right) & \mbox{for $
\overline{\epsilon}^{(1)} \leq 1/2$}
\\ 1 & \mbox{for $ \overline{\epsilon}^{(1)}  \geq 1/2$}
\end{array} \; .
\right.
\end{equation}

\section{Maximizing $p_c^{(1)}$ for corrected errors}
\label{corrected}
We start from equation (\ref{pccorrected}) and use the Cauchy
inequality in a similar way as in appendix \ref{optimisation}. We
obtain the bound
%\widetext
\begin{eqnarray}
p_c^{(1)} &\leq& 1 - \frac{\left({\bf A}_{00} {\bf A}_{10} \right)^2
+\left({\bf A}_{00} {\bf A}_{11} \right)^2 +\left({\bf A}_{01} {\bf
A}_{10} \right)^2 +\left({\bf A}_{01} {\bf A}_{11} \right)^2 }{\left(
\left|{\bf A}_{00}\right|^2+\left|{\bf A}_{01}\right|^2+\left|{\bf
A}_{10}\right|^2+\left|{\bf A}_{11}\right|^2\right)^2} \\ & &
-\frac{\left({\bf \tilde{B}}_{00} {\bf \tilde{B}}_{10} \right)^2
+\left({\bf \tilde{B}}_{00} {\bf \tilde{B}}_{11} \right)^2 +\left({\bf
\tilde{B}}_{01} {\bf \tilde{B}}_{10} \right)^2 +\left({\bf
\tilde{B}}_{01} {\bf \tilde{B}}_{11} \right)^2 }{\left( \left|{\bf
\tilde{B}}_{00}\right|^2+\left|{\bf
\tilde{B}}_{01}\right|^2+\left|{\bf
\tilde{B}}_{10}\right|^2+\left|{\bf \tilde{B}}_{11}\right|^2\right)^2}
\; . \nonumber
\end{eqnarray}
%\narrowtext
%\begin{multicols}{2}
Next we introduce the angles $\varphi_{00}^{11}, \varphi_{00}^{10},
\varphi_{01}^{10}$ between the corresponding vectors ${\bf
A}_{00},{\bf A}_{10},{\bf A}_{01},{\bf A}_{11}$, make use of the
relations (\ref{vectorrelation}) and (\ref{dotproductrelation}), use
the symmetry argument as in appendix \ref{optimisation} and find after
some transformation the set of equations
\begin{eqnarray}
\lefteqn{p_c^{(1)}  \leq \frac{3}{4}}\\
& &  + \frac{\left| {\bf A}_{00} \right|^4
(1-3 \cos^2 \varphi_{00}^{11}) + \left| {\bf A}_{01} \right|^4 ( 1-3
\cos^2 \varphi_{01}^{10})}{8 (\left| {\bf A}_{00} \right|^2 + \left|
{\bf A}_{01} \right|^2)^2} \nonumber \\
 & & + \left| {\bf A}_{00} \right|^2
\left| {\bf A}_{01} \right|^2 \frac{ 3 + \cos \varphi_{00}^{11} \cos
\varphi_{01}^{10} - 2 \cos^2 \varphi_{00}^{10}}{4 (\left| {\bf A}_{00}
\right|^2 + \left| {\bf A}_{01} \right|^2)^2} \nonumber \\
\overline{\epsilon}^{(1)} & = & \frac{ \left| {\bf A}_{00} \right|^2 (
1 - \cos \varphi_{00}^{11}) + \left| {\bf A}_{01} \right|^2 ( 3 - \cos
\varphi_{01}^{10})}{4(\left| {\bf A}_{00} \right|^2 + \left| {\bf
A}_{01} \right|^2)}
\end{eqnarray}
The first observation is that it is optimal to choose $\cos
\varphi_{00}^{10} = 0$ since this choice optimizes  $p_c^{(1)}$ while
it leaves $\overline{\epsilon}^{(1)}$ unchanged. The second
observation is that the choice of
\begin{equation}
\label{symmcond}
\left| {\bf A}_{00} \right|^2 \cos \varphi_{00}^{11} = \left| {\bf
A}_{01} \right|^2 \cos \varphi_{01}^{10}
\end{equation}
 within the subspace defined by $$\left| {\bf A}_{00} \right|^2 \cos
\varphi_{00}^{11} + \left| {\bf A}_{01} \right|^2 \cos
\varphi_{01}^{10} = const$$ and fixed values of $\left| {\bf A}_{00}
\right|$ and $\left| {\bf A}_{01} \right|$ is optimal if this choice
is possible.  In this case we are left with the equations
\begin{eqnarray}
\lefteqn{p_c^{(1)} \leq \frac{3}{4}} \\ & & + \frac{\left| {\bf
A}_{00} \right|^4 (1-4 \cos^2 \varphi_{00}^{11}) + \left| {\bf A}_{01}
\right|^4 + 6\left| {\bf A}_{00} \right|^2 \left| {\bf A}_{01}
\right|^2}{8 (\left| {\bf A}_{00} \right|^2 + \left| {\bf A}_{01}
\right|^2)^2}\nonumber \\ \overline{\epsilon}^{(1)} & = & \frac{
\left| {\bf A}_{00} \right|^2 ( 1 - 2 \cos \varphi_{00}^{11}) + 3
\left| {\bf A}_{01} \right|^2 }{4(\left| {\bf A}_{00} \right|^2
+\left| {\bf A}_{01} \right|^2)} \; .
\end{eqnarray}
At the end of a short maximization calculation we find a solution
consistent with symmetry condition (\ref{symmcond}) for $\frac{1}{4}
\leq \overline{\epsilon}^{(1)} \leq \frac{1}{2}$. It is given by
\begin{equation}
p_c^{(1)} \leq \frac{3}{4} + \overline{\epsilon}^{(1)} -
\left(\overline{\epsilon}^{(1)}\right)^2 \; .
\end{equation}
This maximum is obtained by choosing the values $ \cos
\varphi_{00}^{11} = \frac{1-2\overline{\epsilon}^{(1)}
}{2(1-\overline{\epsilon}^{(1)})} $ and $\left| {\bf A}_{01} \right| =
\left| {\bf A}_{00} \right| \sqrt{\frac{
\overline{\epsilon}^{(1)}}{1-\overline{\epsilon}^{(1)}}}$. The
symmetry condition (\ref{symmcond}) then gives $ \cos
\varphi_{01}^{10} =
\frac{1-2\overline{\epsilon}^{(1)}}{2\overline{\epsilon}^{(1)}}$ which
limits the range of validity to $\frac{1}{4} \leq
\overline{\epsilon}^{(1)}$.  For $\frac{1}{4} \geq
\overline{\epsilon}^{(1)}$ we find the optimal solution by selecting $
\cos \varphi_{01}^{10} =1$. A short maximization calculation then
gives the bound
\begin{equation}
p_c^{(1)} \leq \frac{1}{2} + 3 \overline{\epsilon}^{(1)} - 5
\left(\overline{\epsilon}^{(1)}\right)^2
\end{equation}
for the choice of parameters $ \cos \varphi_{00}^{11} =
\frac{1-3\overline{\epsilon}^{(1)} }{1-\overline{\epsilon}^{(1)}} $
and $\left| {\bf A}_{01} \right| = \left| {\bf A}_{00} \right|
\sqrt{\frac{
\overline{\epsilon}^{(1)}}{1-\overline{\epsilon}^{(1)}}}$.

\section{ Maximizing $p_c^{(1)}$ for corrected errors with leaked
error positions}
\label{correctedleaked}
 We apply Cauchy inequalities to equation (\ref{pccorrpos}) and use
the vector notations ${\bf A}$, ${\bf \tilde{B}}$, ${\bf C}$, and
${\bf \tilde{D}}$ to find
\begin{eqnarray}
\label{pcpc}
\lefteqn{ p_c^{(1)} \leq 1}\\ & & - \frac{1}{4 p_{\rm sif}}
\frac{\left| {\bf A}_{00} {\bf A}_{11} \right|^2}{|{\bf A}_{00}|^2 +
|{\bf A}_{11}|^2} - \frac{1}{4 p_{\rm sif}} \frac{\left| {\bf C}_{01}
{\bf C}_{10} \right|^2}{|{\bf C}_{01}|^2 + |{\bf C}_{10}|^2}\nonumber
\\ & & - \frac{1}{4 p_{\rm sif}} \frac{\left| {\bf \tilde{B}}_{00}
{\bf \tilde{B}}_{11} \right|^2}{|{\bf \tilde{B}}_{00}|^2 + |{\bf
\tilde{B}}_{11}|^2} - \frac{1}{4 p_{\rm sif}} \frac{\left| {\bf
\tilde{D}}_{01} {\bf \tilde{D}}_{10} \right|^2}{|{\bf
\tilde{D}}_{01}|^2 + |{\bf \tilde{D}}_{10}|^2} \; . \nonumber
\end{eqnarray}
It becomes clear immediately that we can replace ${\bf C}$ by ${\bf
A}$ and ${\bf \tilde{D}}$ by ${\bf \tilde{B}}$ because of relations
similar to (\ref{dotproductrelation}). Similar to the calculations in
appendices \ref{optimisation} and \ref{corrected} we introduce the
angles $\varphi_{00}^{11}, \varphi_{00}^{10}, \varphi_{01}^{10}$ and
use the relations (\ref{vectorrelation}) and
(\ref{dotproductrelation}) and the symmetry argument introduced in
appendix \ref{optimisation} to find the new form of (\ref{pcpc}) as
\widetext
\begin{eqnarray}
p_c^{(1)} & \leq & \frac{3}{4} - \frac{|{\bf A}_{00}|^2 \cos^2
 \varphi_{00}^{11} + |{\bf A}_{01}|^2 \cos^2
 \varphi_{01}^{10}}{4(|{\bf A}_{00}|^2 + |{\bf A}_{01}|^2)} \\ & & +
 \frac{|{\bf A}_{00}|^2 |{\bf A}_{01}|^2}{2(|{\bf A}_{00}|^2 + |{\bf
 A}_{01}|^2)} \left[ \frac{(1+ \cos \varphi_{00}^{11}) (1+ \cos
 \varphi_{01}^{10})}{|{\bf A}_{00}|^2 (1+ \cos \varphi_{00}^{11}) +
 |{\bf A}_{01}|^2 ( 1 + \cos \varphi_{01}^{10})} +\right. \nonumber \\ & &
 \mbox{\hspace{2cm}} \left.  \frac{(1- \cos \varphi_{00}^{11})(1- \cos
 \varphi_{01}^{10})}{|{\bf A}_{00}|^2 (1- \cos \varphi_{00}^{11}) +
 |{\bf A}_{01}|^2 ( 1 - \cos \varphi_{01}^{10})}\right] \nonumber 
\end{eqnarray}
%\narrowtext
while we take from appendix \ref{corrected} the expression for
$\overline{\epsilon}^{(1)}$ as
\begin{equation}
\overline{\epsilon}^{(1)} = \frac{ \left| {\bf A}_{00} \right|^2 ( 1 -
\cos \varphi_{00}^{11}) + \left| {\bf A}_{01} \right|^2 ( 3 - \cos
\varphi_{01}^{10})}{4(\left| {\bf A}_{00} \right|^2 + \left| {\bf
A}_{01} \right|^2)} \; .
\end{equation}

We next perform a variation along the path defined by $\left| {\bf
A}_{00} \right|^2 \cos \varphi_{00}^{11} + \left| {\bf A}_{01}
\right|^2 \cos \varphi_{01}^{10} = const$ and find that $p_c^{(1)}$ is
optimized for the choice $ \cos \varphi_{00}^{11} = \cos
\varphi_{01}^{10}$. An optimization calculation for the remaining
parameters leads to the estimate
\begin{equation}
p_c^{(1)} \leq \frac{1}{2} + 2 \overline{\epsilon}^{(1)} - 2
 \left(\overline{\epsilon}^{(1)}\right)^2
\end{equation}
for a disturbance $\overline{\epsilon}^{(1)} \leq 1/2$. This optimum
is obtained by choosing $\cos \varphi_{00}^{11} = 1-2
\overline{\epsilon}^{(1)}$ and $|{\bf A}_{00}| = |{\bf A}_{01}|
\sqrt{\frac{ 1-
\overline{\epsilon}^{(1)}}{\overline{\epsilon}^{(1)}}}$.


\begin{thebibliography}{10}

\bibitem{bennett84a} C.~H. Bennett and G. Brassard, In {\em
Proceedings of IEEE International Conference on Computers, Systems,
and Signal Processing, Bangalore, India}, (IEEE, New York, 1984)\ pp.\
175--179.

\bibitem{huttner94a} B. Huttner and A.~K. Ekert, J.  Mod. Opt. {\bf
41,} 2455--2466 (1994).

\bibitem{marand95a} C. Marand and P.~T. Townsend,  Opt. Lett. {\bf 20,}
1695--1697 (1995).

\bibitem{zbinden98a} H. Zbinden, N. Gisin, B. Huttner, A. Muller,
J. Cryptol. {\bf 11,} 1--14 (1998).

\bibitem{franson94a}
J.~D. Franson and H. Ilves,  J. Mod. Opt. {\bf 41,} 2391--2396 (1994).

\bibitem{buttler98a} W.~T. Buttler, R.~J. Hughes, P.~G. Kwiat,
G.~G. Luther, G.~L. Morgan, J.~E.  Nordholt, C.~G. Peterson, and
C.~M. Simmons,  Phys. Rev. A
{\bf 57,} 2379--2382 (1998).

\bibitem{bennett95a} C.~H. Bennett, G. Brassard, C. Cr\'epeau, and
U.~M. Maurer, IEEE
Trans. Inf. Theo. {\bf 41,} 1915 (1995).

\bibitem{law97a} C.~K. Law and H.~J. Kimble, 
J. Mod. Opt. {\bf 44,} 2067--2074 (1997).

\bibitem{yuen96a}
H.~P. Yuen, Quantum. Semicl. Opt. {\bf 8,} 939--949 (1996).

\bibitem{huttner95a} {B.~Huttner and N.~Imoto and N.~Gisin and
T.~Mor},  Phys. Rev. A
{\bf 51,} 1863--1869 (1995).

\bibitem{mayers98a}
D. Mayers, Report quant-ph/9802025, (1998).

\bibitem{biham98a} E. Biham, M. Boyer, G. Brassard, J. van~de Graaf,
and T. Mor, Report quant-ph/9801022, (1998).

\bibitem{cachin97a} C. Cachin and U.~M. Maurer,  J. Crypt. {\bf 10,}
97--110 (1997).

\bibitem{brassard93a} G. Brassard and L. Salvail,  In {\em Proceedings of
Eurocrypt '93, held in Lofthus, Norway, 1993},

\bibitem{fuchs97a} C.~A. Fuchs, N. Gisin, R.~B. Griffiths, C.-S. Niu,
and A. Peres, 
Phys. Rev. A {\bf 56,} 1163--1176 (1997).

\bibitem{nl96a} N. L\"utkenhaus,  Phys. Rev. A {\bf 54,} 97 (1996).

\bibitem{hoeffding63a} W. Hoeffding, J. Amer. Stat. Ass {\bf 58,}
13--30 (1963).

\bibitem{dusek} In an earlier version of this paper I omitted the
authentication of this step. I am grateful to Miloslav Du\v{s}ek for
pointed out to me the danger arising from that.

\bibitem{wegman81a}
M.~N. Wegman and J.~L. Carter, J. Comp. Syst. Sci. {\bf 22,} 265--279
  (1981).

\bibitem{davies76a} E.~B. Davies, {\em Quantum Theory of Open Systems}
(Academic Press, London, New York, San Francisco, 1976).

\bibitem{kraus83a} K. Kraus, in {\em States, Effects, and Operations},
No.~190 in {\em Lecture Notes in Physics}, A. B{\"o}hm, J.~D. Dollard,
and W. Wooters, eds., (Springer, Berlin, 1983).

\bibitem{yurke85a} B. Yurke,  Phys. Rev. A {\bf 32,} 311--323 (1985).

\bibitem{nl96b} N. L\"utkenhaus, Ph.D. thesis, University of
Strathclyde, Glasgow, Scotland, 1996.

\bibitem{nl97b} N. L\"utkenhaus and S.~M. Barnett, In {\em Proceedings
of an International Workshop on Quantum Communication, Computing, and
Measurement, held September 25-30 in Shizuoka, Japan}, O. Hirota,
A.~S. Holevo, and C.~M. Caves, eds., (Plenum Press, New York, 1997).

\bibitem{slutsky97a} B. Slutsky, P.~C. Sun, Y. Mazurenko, R. Rao, and
Y. Fainman,  J. Mod. Opt. {\bf 44,} 953--961
(1997).

\bibitem{slutsky98a} B. Slutsky, R. Rao, P.~C. Sun, and Y. Fainman,
Phys. Rev. A {\bf 57,} 2383--2398 (1998).


\end{thebibliography}
\end{document}